\documentclass[12pt, peerreview , onecolumn]{IEEEtran}

\usepackage{tikz}
\usepackage[sumlimits,intlimits]{amsmath}
\usepackage{amsfonts,amssymb}%
\usepackage{bm}%
\usepackage{graphicx,graphics}%
\usepackage{cases}%
\usepackage[noadjust]{cite}%
\usepackage{color}%
\usepackage{url}
\usepackage{verbatim}
\usepackage{balance}
\usepackage{multirow}
\usepackage{multicol}
\usepackage{stfloats}
\usepackage{xspace}
\usepackage{enumerate}
\usepackage{rotating}
\usepackage[tight]{subfigure}
\usepackage{algorithm}
\usepackage{algorithmicx, algpseudocode}

\newtheorem{theorem}{Theorem}

\newcommand{\bk}{\bar{k}}
\newcommand{\bn}{\bar{n}}

\newcommand{\siga}{\sigma^2_{a}}
\newcommand{\sigc}{\sigma^2_{c}}

\newcommand{\ds}{\displaystyle}
\newcommand{\la}{\langle}
\newcommand{\ra}{\rangle}
\newcommand{\clK}{{\cal K}}

\newcommand{\clN}{{\cal N}}
\ifCLASSOPTIONpeerreview

\fi
\begin{document}

\setcounter{page}{1}

\title {Secure and Energy-Efficient Beamforming for Simultaneous Information and Energy Transfer}
\author{Ali A. Nasir,  Hoang D. Tuan, Trung Q. Duong and H. Vincent Poor
\thanks{A.~A.~Nasir is with the Department of Electrical Engineering, King Fahd University of Petroleum and Minerals (KFUPM), Dhahran, Saudi Arabia (Email: anasir@kfupm.edu.sa).}
\thanks{H.~D.~Tuan is with the Faculty of Engineering and Information Technology, University of Technology Sydney, Broadway, NSW 2007, Australia (email: Tuan.Hoang@uts.edu.au).}
\thanks{T.~Q.~Duong is with Queen's University Belfast, Belfast BT7 1NN, UK  (email: trung.q.duong@qub.ac.uk)}%
\thanks{H.~V. Poor is with the Department of Electrical Engineering, Princeton University, Princeton, NJ 08544, USA (e-mail: poor@princeton.edu).}
}

\maketitle


\begin{abstract} Next-generation communication networks will likely involve the energy-efficient transfer of information and energy over the same wireless channel, for which the physical layer will become more vulnerable to cyber attacks by potential multi-antenna eavesdroppers. To address this issue, this paper considers transmit time-switching (TS) mode, in which energy and information signals are transmitted separately in time by the BS. This protocol is not only easy to implement  but also delivers the opportunity of multi-purpose beamforming,  in which energy beamformers during wireless power transfer are useful in jamming the eavesdropper. In the presence of imperfect channel estimation and multi-antenna eavesdroppers, the energy and information beamformers and the transmit TS ratio are jointly optimized to maximize the worst-case user secrecy rate subject to UEs’ harvested energy thresholds and a BS transmit power budget.  New robust path-following algorithms, which involve one simple convex quadratic program at each iteration are proposed for computational solutions of this difficult optimization problem and also  the problem of secure energy efficiency maximization. The latter is further complex due to additional optimization variables appearing in the denominator of the secrecy rate function. Numerical results confirm that the performance of the proposed computational solutions is robust against the channel uncertainties.​
\end{abstract}

\vspace{1cm}

\section{Introduction} \label{sec:int}

Next-generation communication networks offers the potential to transfer information and energy through the same wireless communication channel, where energy constrained users (UEs) would be able to not only receive information but also harvest energy \cite{Lu-14-A,Ding-13-A,Krikidis-14-MCOM-A}. {\color{black}The information transfer generally aims at high signal-to-interference-plus-noise-ratio (SINR) while the energy transfer aims at a high-power ambient
signal \cite{Xu-14-Sep-A,Zheng-15-Feb-A}}.  In early developments, information and energy are {\color{black}excited} to be transferred simultaneously (at the same time) by the same signalling. To realize both wireless energy harvesting (EH) and information decoding (ID), the user's receivers need to split the received signal for EH and ID either by power splitting (PS) or time switching (TS) {\color{black}\cite{Zhang-12-A,Nasir-13-A}}. Our recent result in \cite{Nasir-16-TCOM-A} shows that such protocol, particularly the PS approach at the receiver, is not only complicated and inefficient for practical implementation, but also not necessary. It is {\color{black}much more efficient} to transfer information and energy separately {\color{black}and} the users's receivers do not need any sophisticated device.

Wireless power transfer is more viable in sensor-networks or in dense small-cell deployment where there is closer proximity between BS and UEs. Such densification of wireless network make the wireless devices more vulnerable to malicious cyber attacks than ever
\cite{Fragkiadakis-13-First-A,Hetal15}. The eavesdropper can be more powerful as equipped by multi-antenna and in favorable channel condition. The information intended to users of less favorable channel condition can be vulnerably leaked. Physical layer security aims to secure data transmissions in such networks \cite{P12,LT10,BB11}. Many recent works considered looking into the beamforming design problem to maximize secrecy rate under the BS transmit power budget \cite{Mukherjee-11-Jan-A,Li-13-May-A,Zhao-15-Sep-A,Chu-15-A}. Beamforming requires the knowledge of downlink channels to the UEs, which can be obtained via channel estimation. Due to channel estimation errors in practical systems, the BS cannot expect perfect channel knowledge, which demands for robust beamforming design in the presence of channel uncertainties \cite{Mukherjee-11-Jan-A,Zhao-15-Sep-A}. Adding wireless energy harvesting (EH) feature due to its viability in dense small-cell deployment introduces another EH constraint in secrecy rate optimization problem \cite{Nasir-16-TSP-A}. As mentioned above, physical layer security {\color{black}becomes more} relevant in wireless {\color{black}information and} power transfer systems.

Robust beamforming design  in the presence of channel uncertainties with the same objective of secrecy rate maximization under receiver EH thresholds in addition to BS transmit power budget was recently considered in {\color{black}\cite{Zhang-15-Mar-A,Zhu-16-TWC-A,Feng-15-A,Zhang-16-TWC-A}}. Some of these works assume either only EH feature or only ID capability at the UEs \cite{Feng-15-A,Zhu-16-TWC-A}, so there were no
PS or TS based simultaneous wireless information and power transfer (SWIPT) receivers. Assuming PS-based SWIPT receivers, secrecy rate maximization was studied in \cite{Zhang-15-Mar-A,Zhang-16-TWC-A}.
{\color{black}At griding points of normal rates},  these works employ semi-definite programming {\color{black}and alternating optimization}, where rank-one constraints have to be dropped and computationally complex matrices have to be optimized.  Randomization has to be employed to achieve feasible beamforming vectors \cite{Zhang-16-TWC-A}.
As already pointed out by \cite{Phan-12-A} while ago, such randomization approach is not quite efficient. Moreover, with the
existing PS approach, it is well known that it is not practically easy to implement variable range power splitter and also, one can not jam the eavesdropper without transmitting artificial noise \cite{Nasir-16-TCOM-A}. In contrast, as shown in the
present paper, our recently proposed transmit TS approach \cite{Nasir-16-TCOM-A} does not require to transmit extra artificial noise thanks to the fact that power-bearing signal sent during EH time can be simultaneously used to jam the eavesdropper.

{\color{black}Meanwhile, energy efficiency (EE) in terms of bits per Joule per Hertz is also a very
important figure-of-merit in assessing the practicability of next communication networks and beyond (see e.g. {\color{black}\cite{Caval14,Ietal14,Betal16,Ng-12-Jul-TVT-A,Chen-13-COMML-A,
Zappone-16-Dec-A}}), where the Dinkelbach-type algorithm \cite{D67} of fractional programming
is the main tool for obtaining computational solutions (see e.g. \cite{ZJ15,Zetal16} and references therein).} In the presence of eavesdroppers, secrecy energy efficiency (SEE) maximization has been studied recently in \cite{Zhao-15-TVT-A,Vu-16-COMML-A}. However, the approach to treat SEE in \cite{Zhao-15-TVT-A,Vu-16-COMML-A} is based on costly  beamformers, which completely cancel the multi-user interference and wiretapped signal at the eavesdroppers. Moving step ahead, energy harvesting {\color{black}brings} in conflicting requirements form the viewpoint of EE, as it requires a stronger {\color{black}transmit power}. The problem of energy efficiency maximization in SWIPT systems has been recently studied in \cite{Guo-15-SECON-P,Vu-12-SPL-A,Leng-15-ICNC-P}. However, either the authors don't consider simultaneous EH and ID capability \cite{Leng-15-ICNC-P} or assume PS based receiver \cite{Guo-15-SECON-P,Vu-12-SPL-A}. To the best of our knowledge, computational solution for robust beamforming design to achieve secrecy rate and SEE optimization, particularly assuming practical TS-based wireless EH systems, is still an open problem. {\color{black}The SEE objective is not a ratio of concave and convex functions, for which
 the Dinkelbach's algorithm based approach is very inefficient}.

{\color{black}The subject of this paper is  a multicell network, where the UEs in each cell are
 divided into two groups depending upon their distance from the serving BS.}
 The one closer to the BS take the advantage of higher received power to perform wireless EH in addition to ID while the far-away users only conduct ID. We consider imperfect channel state information (CSI) case where the BSs have imperfect channel knowledge about UEs and eavesdroppers. We implement transmit TS approach \cite{Nasir-16-TCOM-A} where BS transmits information and energy separately in different time portions and the energy beamformers can be exploited to jam the eavesdroppers.\footnote{Though we propose transmit TS approach to solve max-min rate and power minimization problems in \cite{Nasir-16-TCOM-A}, however, the extension of those developed algorithms to solve robust secrecy rate and energy efficiency maximization problem in the presence of eavesdroppers and channel estimation errors (as will be detailed shortly in this paper) is highly non-trivial.} In the presence of channel uncertainties, we formulate the worst-case based robust secrecy rate optimization problem. We solve for joint optimization of information and energy beamforming vectors with the transmit TS ratio, that could maximize the minimum secrecy rate among all users, while ensuring EH constraints for near-by users and transmit power constraints at the BSs.
 {\color{black}The problem is  very difficult computationally  due many challenging constraints, for
 which a path-following algorithm is developed for its computational solution. The algorithm does not require rank-constrained optimization and converges quite quickly in few iterations.  Through extensive simulation, the achieved secrecy rate
 is shown to be close to the  normal rate that excludes the presence of eavesdroppers. } Furthermore, our numerical results confirm that performance of the proposed algorithm is close to that of the perfect channel knowledge case. In addition, the proposed algorithm not only outperforms the existing algorithm that models power-splitting (PS) based receiver but also the proposed transmit TS based model is implementation-wise quite simple than the PS-based model. In the end, we extend our development to solve and analyze robust SEE maximization problem, which is further complex due to additional function of optimization variables in the denominator of secrecy rate function.

%
%
%
%
%
%
%
%
%
%

{\color{black}The paper is organized as follows. Section II presents the problem formulation for maximizing the worst-case user
secrecy rate  and its challenges, whereas Section III develops its computational solution. Section IV proposes a computational
solution for the EE maximization. Section V evaluates the performance of our proposed algorithms by numerical examples. Finally, Section VI concludes the paper.}

\textit{Notation}. We use $\Re\{\cdot\}$ operator to denote the
{\color{black}real part of its argument}, $\nabla$ operator to denote the first-order differential operator, and $\| \mathbf{x} \|$ and $\| \mathbf{X} \|_F$ to denote the Euclidean and Frobenius norm of a vector $\mathbf{x}$ and matrix $\mathbf{X}$, respectively. Also, we define $\la \mathbf{x},\mathbf{y} \ra \triangleq \mathbf{x}^H \mathbf{y}$.

\begin{figure}[t]
    \centering
    \includegraphics[width=0.65 \textwidth]{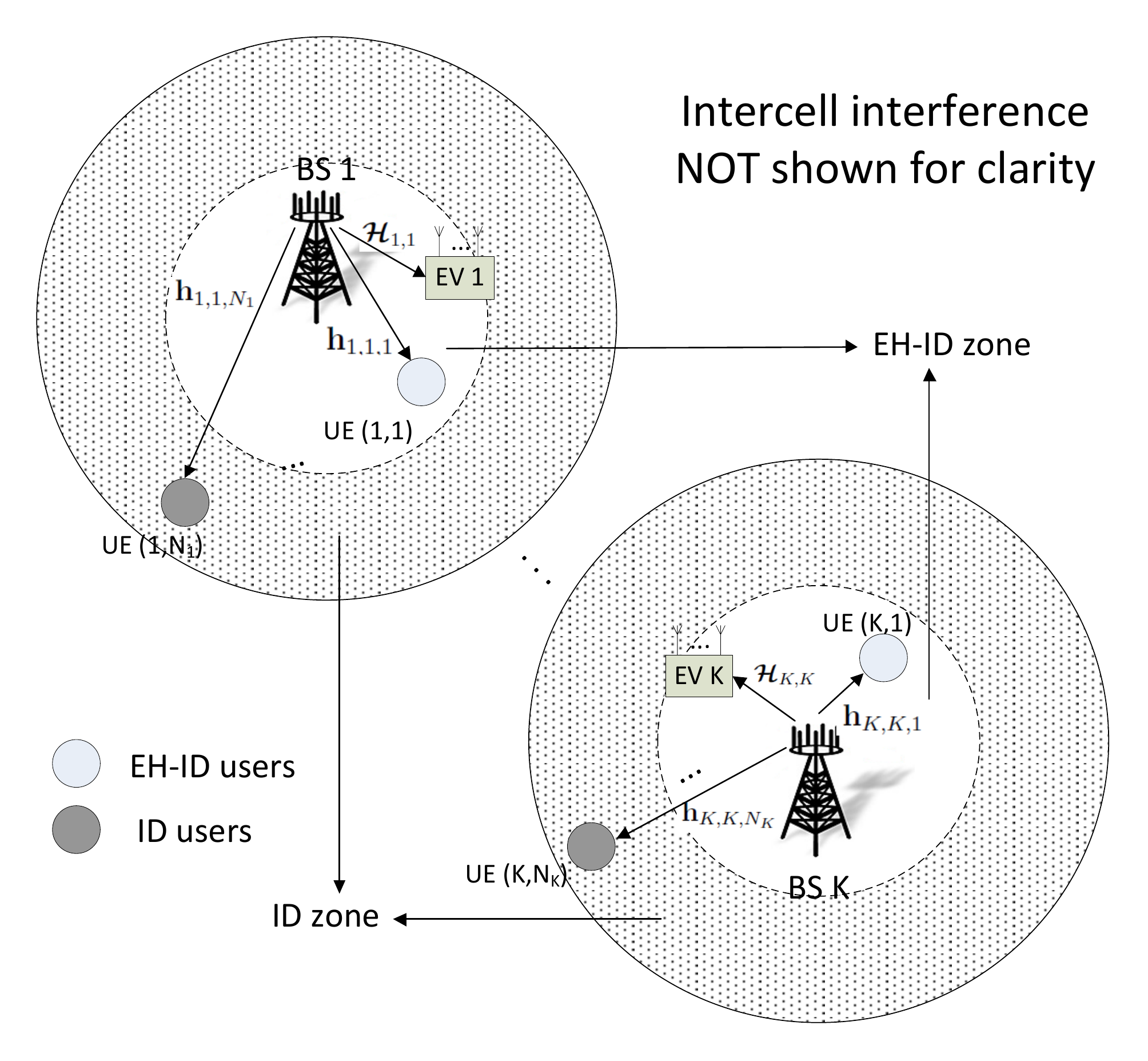} 
  \caption{{\color{black}Downlink multiuser multicell interference scenario in a dense network consisting of $K$ small cells. For clarity, the intercell interference channels are not shown, however, the interference occurs in all $K$ cells.}}
    \label{fig:sys_mod}
\end{figure}

\section{System Model and Problem Formulation}

{\color{black}Consider} a multicell network consisting of $K$ small cells labeled by
$k\in\mathcal{K}\triangleq\{1,\dots,K\}$.
{\color{black}As shown in Fig. \ref{fig:sys_mod}}, in each cell $k$, a multi-antenna BS $k$ with $M$ antennas communicates with $N_k$ single-antenna users (UEs)
$(k,n)$,  $n \in \mathcal{N}_k \triangleq \{1,\dots,N_k\}$
 over the same bandwidth. We divide the users in each cell $k$ into two zones, such that there are $N_{1,k}$ users located nearby serving BS $k$ in zone-1 and $N_{2,k}$ users are located far from the BS $k$ in zone-2, where $N_k = N_{1,k} + N_{2,k}$. By UE $(k,n_1)$ and UE $(k,n_2)$, we mean UE $n_1 \in \mathcal{N}_{1,k} \triangleq \{1,\dots,N_{1,k}\}$ in zone-1 and UE $n_2 \in \mathcal{N}_{2,k} \triangleq \{N_{1,k}+1,\dots,N_{k}\}$ in zone-2 of cell $k$, respectively. {\color{black}Moreover, as shown in Fig. \ref{fig:sys_mod}, we assume that for UEs $(k,n)$ of cell $k$, there is {\color{black}a} single eavesdropper $k$ with $N_\text{ev}$ antennas in  zone-1, who {\color{black}eavesdrops} upon the signals
  intended  for UEs $(k,n)$.}

BSs intend to transfer energy to only their zone-1 users since they are located sufficiently {\color{black}near}
to their serving BSs and are able to practically harvest energy. Information is transmitted to both zone-1 and zone-2 users.
{\color{black}Denote by $\mathbf{x}^E_{k,n_1} \in \mathbb{C}^{M \times 1}$ and $\mathbf{x}^I_{k,n} \in \mathbb{C}^{M \times 1}$
the EH beamforming vector and ID beamforming vectors by BS $k$ for its UE ($k,n_1$) and UE ($k,n$), respectively.}
{\color{black} The channel $\widetilde{\mathbf{h}}_{\bar{k},k,n} \in \mathbb{C}^{M \times 1}$ between BS $\bar{k}$ and UE ($k,n$)
is assumed to be frequency flat fading, which counts the effects of both large-scale pathloss and small-scale fading.
Denote $s^E_{k,n_1}$ and $s^I_{k,n}$  as the energy signal and information signal  intended for UE ($k,n_1$) and UE ($k,n$)
by BS $k$,  with $\mathbb{E}\{ | s^E_{k,n_1} |^2 \} = \mathbb{E}\{ | s^I_{k,n} |^2 \}=1$.} Let $0<\eta<1$ be the time splitting for transferring
 energy and information to UE. The baseband signal received by UE ($k,n_1$) for EH is
\begin{align}\label{eq:y_kn}
   y^E_{k,n_1}  &= \sum_{\bk\in\mathcal{K}} \widetilde{\mathbf{h}}_{\bar{k},k,n_1}^H \sum_{\bn\in\mathcal{N}_{1,\bk}} \mathbf{x}^E_{\bar{k},\bar{n}} s^E_{\bar{k},\bar{n}} +  z_{k,n_1}^a,
\end{align}
where $z_{k,n_1}^a \sim \mathcal{CN}(0,\siga)$ is the additive white complex Gaussian noise, with zero-mean and variance $\siga$, at the receiver of UE ($k,n$). Using \eqref{eq:y_kn} {\color{black}and assuming a linear EH model}\footnote{\color{black}The recently studied non-linear EH model and waveform design for efficient wireless power transfer \cite{Clerckx-16-Dec-A,Clerckx-17-A,Zeng-17-May-A} is beyond the scope of this work, but could be incorporated in future research.}, the harvested energy by the UE $(k,n_1)$ can be written as
\begin{align}\label{harvest}
   E_{k,n_1}(\mathbf{x}^E,\eta) &\triangleq \zeta_{k,n_1}\eta
   p_{k,n_1}(\mathbf{x}^E),
\end{align}
where
\begin{align}\label{pkn1}
p_{k,n_1}(\mathbf{x}^E)\triangleq  \sum_{\bk\in\mathcal{K}} \sum_{\bn\in\mathcal{N}_{1,\bk}} | \widetilde{\mathbf{h}}_{\bk,k,n_1}^H \mathbf{x}^E_{\bk,\bn}  |^2  + \siga,
\end{align}
and $\zeta_{k,n_1} \in (0,1)$ is the energy conversion efficiency for the EH receiver. {\color{black}Here, we assume a common TS ratio $\eta$ for all BSs, $k \in \mathcal{K}$, where near-by users harvest energy through wireless signals not only from the serving BSs but also from the neighboring BSs.} Note that the harvested and stored energy $E_{k,n_1}$ may be used later for different power constrained operations at UE $(k,n_1)$, e.g., assisting uplink data transmission to the BS or performing downlink information processing. There and after $\mathbf{x}^E \triangleq [\mathbf{x}^E_{k,n_1}]_{k\in{\cal K}, n_1\in {\cal N}_{1,k}}$. The received signal by UE ($k,n$) for ID is
\begin{equation}\label{eq:y_kn_explicit}
 y^I_{k,n}= \widetilde{\mathbf{h}}_{k,k,n}^H \mathbf{x}^I_{k,n} s^I_{k,n} + \widetilde{\mathbf{h}}_{k,k,n}^H \sum_{\bn\in\mathcal{N}_k\setminus{\{n\}}} \mathbf{x}^I_{k,\bn} s^I_{k,\bn} + \sum_{\bar{k}\in\mathcal{K}\setminus\{k\}} \widetilde{\mathbf{h}}_{\bk,k,n}^H \sum_{\bn\in\mathcal{N}_{\bk}}\mathbf{x}^I_{\bk,\bn} s^I_{\bk,\bn} +  z_{k,n}^a,
\end{equation}
where {\color{black}its first term represents the desired signal,} {\color{black}while the second and third terms are the intracell interference and intercell interference.} The BSs are assumed to perform channel estimation to acquire channel knowledge $\mathbf{h}_{\bk,k,n}$ and the channel state information (CSI) errors are bounded by the uncertainty $\epsilon_{\bk,k,n}$ as follows
\cite{HS12,RTN13}:
 \begin{equation}\label{un0}
 \rho(\widetilde{\mathbf{h}}_{\bk,k,n}\widetilde{\mathbf{h}}^H_{\bk,k,n}-
 \mathbf{h}_{\bk,k,n} \mathbf{h}_{\bk,k,n}^H )\leq \epsilon_{\bk,k,n},
 \end{equation}
where $\rho(A)$ is called the spectral radius of matrix $A$: $\rho(A)=\max_{i}|\lambda_i(A)|$ with its eigenvalues $\lambda_i(A)$, and the channel uncertainties $\epsilon_{\bk,k,n}$ are given by
\begin{align}
\epsilon_{\bk,k,n} = \begin{cases}
\epsilon_0 \| \mathbf{h}_{\bk,k,n} \|^2,& \ \ \ \  k \ne \bk \\
\epsilon_1 \| \mathbf{h}_{\bk,k,n} \|^2,& \ \ \ \ k = \bk,
\end{cases}
\end{align}
where $\epsilon_0$ and $\epsilon_1$ are the normalized uncertainty levels related to neighboring cells' UEs and the serving cells' UEs, respectively.\footnote{We have introduced two different uncertainty levels because later we will show in Section \ref{sec:sim} that secrecy rate is more sensitive to the estimation errors of serving users' channels compared to that of the neighboring users' channels.} Note that (\ref{un0}) covers all uncertainty structures
\cite{RTN13}. Thus, incorporating the channel uncertainties, the worst-case information rate decoded by UE $(k,n)$ is given by \cite{RTN13}
\begin{align}
 (1-\eta) \log_2(1+\text{SINR-UE}_{k,n}) \triangleq (1-\eta)\log_2 \left(1+\ds\frac{|\mathbf{h}_{k,k,n}^H \mathbf{x}^I_{k,n} |^2 - \epsilon_{k,k,n} \| \mathbf{x}_{k,n}^I  \|^2 }{\varphi_{k,n}(\mathbf{x}^I)}\right)\label{sinr}
\end{align}
where $\mathbf{x}^I \triangleq [\mathbf{x}^I_{k,n}]_{k\in{\cal K}, n\in {\cal N}_k}$ and
\begin{align}\label{eq:varphi}
\varphi_{k,n}(\mathbf{x}^I) &\triangleq
\underbrace{\sum_{\bn\in\mathcal{N}_k\setminus\{n\}} |\mathbf{h}_{k,k,n}^H \mathbf{x}^I_{k,\bn} |^2}_\text{intracell interference} + \underbrace{\sum_{\bk\in\mathcal{K}\setminus\{k\}} \sum_{\bn\in\mathcal{N}_{\bk}} |\mathbf{h}_{\bk,k,n}^H \mathbf{x}^I_{\bk,\bn} |^2}_\text{intercell interference}  \notag \\ & \hspace{3cm}  +   \sum_{\bn\in\mathcal{N}_k\setminus\{n\}}  \epsilon_{k,k,n} \| \mathbf{x}^I_{k,\bn} \|^2  +  \sum_{\bk\in\mathcal{K}\setminus\{k\}} \sum_{\bn\in\mathcal{N}_{\bk}} \epsilon_{\bk,k,n} \| \mathbf{x}^I_{\bk,\bn} \|^2 + \siga .
\end{align}

A multi-antenna eavesdropper with $N_\text{ev}$ antennas tries to eavesdrop the intended signals for the UE $(k,n)$. The signal received at the EV $k$ is {\color{black}composed of the signal received
during time fraction $\eta$}, denoted by $\mathbf{y}_{k}^E \in \mathbb{C}^{N_\text{ev} \times 1}$ and given by
\begin{eqnarray}
 \mathbf{y}_{k}^E&=& \sum_{\bar{k}\in\mathcal{K}} \widetilde{\boldsymbol{\mathcal{H}}}^H_{\bar{k},k}   \sum_{\bn\in\mathcal{N}_{1,\bk}}  \mathbf{x}^E_{\bk,\bn} s^E_{\bk,\bn} +  \boldsymbol{z}_{k}^a,
 \notag
\end{eqnarray}
and the signal received at the EV $k$ during  time fraction $1-\eta$, denoted by $\mathbf{y}_{k}^I \in \mathbb{C}^{N_\text{ev} \times 1}$ given by
\begin{eqnarray}
 \mathbf{y}_{k}^I&=& \sum_{\bar{k}\in\mathcal{K}} \widetilde{\boldsymbol{\mathcal{H}}}^H_{\bar{k},k}   \sum_{\bn\in\mathcal{N}_{\bk}}
 \mathbf{x}^I_{\bk,\bn} s^I_{\bk,\bn} +  \boldsymbol{z}_{k}^a,
 \notag
\end{eqnarray}
where $\widetilde{\boldsymbol{\mathcal{H}}}_{\bk,k}$ is the wiretap channel matrix of size $M \times N_\text{ev}$ between BS $\bk$ and UE $k$ and $\boldsymbol{z}_{k}^a\in\mathbb{C}^{N_{ev}}\left(0,\sigma_a^2 I_{N_\text{ev}} \right)$ is noise \cite{Detal10,LPW11,Detal12,Hetal15}. Since the eavesdropper is not aware of the time switching factor $\eta$,
{\color{black}$\mathbf{y}_k^E$ is considered as an additional noise to jam the eavesdropper.}
{\color{black}Therefore, the  noise power at EV $k$ in decoding $s^I_{k,n}$ is given by
\begin{eqnarray}
\eta  \sum_{\bar{k}\in\mathcal{K}}   \sum_{\bn\in\mathcal{N}_{1,\bk}}  \| \widetilde{\boldsymbol{\mathcal{H}}}^H_{\bar{k},k}   \mathbf{x}^E_{\bk,\bn} \|^2  +  (1-\eta) (\sum_{\bar{k}\in\mathcal{K}}  \sum_{\bn\in\mathcal{N}_{\bk}}
 \| \widetilde{\boldsymbol{\mathcal{H}}}^H_{\bar{k},k}   \mathbf{x}^I_{\bk,\bn} \|^2 -
 \| \widetilde{\boldsymbol{\mathcal{H}}}^H_{k,k} \mathbf{x}^I_{k,n} \|^2  ) +  N_{\text{ev}} \sigma_a^2.
 \label{eq:y_kn_prime}
\end{eqnarray}
}
 We assume that the wiretap channel state information
 $\boldsymbol{\mathcal{H}}_{\bk,k}$ is available through channel estimation subject to some
 uncertainty \cite{HS12,RTN13}
 \begin{equation}\label{un1}
 \rho(\widetilde{\boldsymbol{\mathcal{H}}}_{\bk,k}\widetilde{\boldsymbol{\mathcal{H}}}^H_{\bk,k}-
 \boldsymbol{\mathcal{H}}_{\bk,k}\boldsymbol{\mathcal{H}}^H_{\bk,k})\leq \epsilon_{\bk,k}, \ \forall \bk,k \in \mathcal{K},
 \end{equation}
where $\epsilon_{\bk,k} = \epsilon_0 \| \boldsymbol{\mathcal{H}}_{\bk,k} \|_F^2$ and $\epsilon_0$  is the normalized uncertainty level for the channels between BSs and the eavesdroppers. Therefore, the worst received SINR at the EV $k$, corresponding to the signal targeted for the UE $(k,n)$, is given by \cite{RTN13}
\begin{align}
    \text{SINR-EV}_{k,n} \triangleq \ds\frac{\|\boldsymbol{\mathcal{H}}_{k,k}^H \mathbf{x}^I_{k,n} \|^2 +\epsilon_{k,k} \|\mathbf{x}^I_{k,n}\|^2 }{ q_{k,n}(\mathbf{x},\eta) }.\label{sinr_ev}
\end{align}
where
\begin{align} \label{eq:qk}
q_{k,n}(\mathbf{x},\eta)& \triangleq \ds\frac{\eta}{(1-\eta)}\left( \ds\sum_{\bk\in\mathcal{K}} \sum_{\bn\in\mathcal{N}_{1,k}} \|\boldsymbol{\mathcal{H}}_{\bk,k}^H \mathbf{x}^E_{\bk,\bn} \|^2
{\color{black} - }  \ds\sum_{\bk\in\mathcal{K}} \sum_{\bn\in\mathcal{N}_{1,k}} \epsilon_{\bk,k} \|\mathbf{x}^E_{\bk,\bn} \|^2 \right) \notag \\
 &+ \ds\sum_{\bn\in\mathcal{N}_{k} \setminus \{n\}} \|\boldsymbol{\mathcal{H}}_{k,k}^H \mathbf{x}^I_{k,\bn} \|^2 +\ds\sum_{\bk\in\mathcal{K}\setminus\{k\}} \sum_{\bn\in\mathcal{N}_{\bk}} \|\boldsymbol{\mathcal{H}}_{\bk,k}^H \mathbf{x}^I_{\bk,\bn} \|^2 \notag \\
    & {\color{black} - } \left(\ds\sum_{\bn\in\mathcal{N}_{k} \setminus \{n\}} \epsilon_{k,k} \|\mathbf{x}^I_{k,\bn} \|^2+ \ds\sum_{\bk\in\mathcal{K}\setminus\{k\}} \sum_{\bn\in\mathcal{N}_{\bk}}   \epsilon_{\bk,k}  \|\mathbf{x}^I_{\bk,\bn} \|^2 \right)  + N_\text{ev}\siga/(1-\eta),
 \end{align}
where $\mathbf{x} \triangleq \left[  \mathbf{x}^E ; \mathbf{x}^I \right]$.

{\color{black} The main attractive feature in (\ref{sinr_ev})-(\ref{eq:qk}) is that the EH signals contribute very much
to the denominator of the SINR (\ref{sinr_ev}) at EV $k$, i.e. they are also used in jamming the EV $k$.}
 The secrecy rate expression for UE $(k,n)$ in nat/sec/Hz is given as \cite{Letal07}
\begin{eqnarray}
f_{k,n}(\mathbf{x},\eta)&=&(1-\eta)\ln(1+\text{SINR-UE}_{k,n})-\ln(1+\text{SINR-EV}_{k,n} )\notag \\
&=&(1-\eta)\ln \left(1+\ds\frac{|\mathbf{h}_{k,k,n}^H \mathbf{x}^I_{k,n} |^2 - \epsilon_{k,k,n} \| \mathbf{x}_{k,n}^I  \|^2}{\varphi_{k,n}(\mathbf{x}^I)} \right)
-\ln \left(1+\ds\frac{ \|\boldsymbol{\mathcal{H}}_{k,k}^H \mathbf{x}^I_{k,n} \|^2  +\epsilon_{k,k} \|\mathbf{x}^I_{k,n}\|^2 }{q_{k,n}(\mathbf{x},\eta) } \right),\notag \\ &=& (1-\eta)f_{k,n}^1(\mathbf{x}^I) - f_{k,n}^2(\mathbf{x},\eta)\label{srate2}
\end{eqnarray}
where
\[
f_{k,n}^1(\mathbf{x}^I) \triangleq \ln \left(1+\ds\frac{|\mathbf{h}_{k,k,n}^H \mathbf{x}^I_{k,n} |^2 - \epsilon_{k,k,n} \| \mathbf{x}_{k,n}^I  \|^2}{\varphi_{k,n}(\mathbf{x}^I)} \right)
\]
and
\[
f_{k,n}^2(\mathbf{x},\eta) \triangleq \ln \left(1+\ds\frac{ \|\boldsymbol{\mathcal{H}}_{k,k}^H \mathbf{x}^I_{k,n} \|^2 +\epsilon_{k,k} \|\mathbf{x}^I_{k,n}\|^2 }{q_{k,n}(\mathbf{x},\eta) } \right).
\]
The corresponding rate can be calculated in bits/sec/Hz units by evaluating $\frac{f_{k,n}(\mathbf{x},\eta)}{\ln 2}$.

At first, we aim to jointly optimize the transmit information and energy beamforming vectors, $\mathbf{x}^E_{k,n_1}$ and $\mathbf{x}^I_{k,n}$, respectively, and the TS ratio $\eta$ to maximize the minimum (user with worst channel conditions) secrecy rate
\begin{subequations} \label{eq:P1}
\begin{align}
  \ds  \max_{\overset{\mathbf{x}^E_{k,n_1},\mathbf{x}^I_{k,n} \in\mathbb{C}^{M\times 1}}{\eta\in(0,1)}} & F(\mathbf{x},\eta) \triangleq
  \ds\min_{k\in\mathcal{K}, n\in\mathcal{N}_k} f_{k,n}(\mathbf{x},\eta) =
  \ds\min_{k\in\mathcal{K}, n\in\mathcal{N}_k} \left[ \ds (1-\eta)f_{k,n}^1(\mathbf{x}^I) -
  f_{k,n}^2(\mathbf{x},\eta) \right]\quad
   \label{eq:Omm1}\\
    \text{s.t.} \quad &g_k(\mathbf{x}_k)\triangleq \ds\eta\sum_{n_1\in\mathcal{N}_{1,k}} \| \mathbf{x}^E_{k,n_1} \|^2 +
    (1-\eta)\sum_{n\in\mathcal{N}_k} \| \mathbf{x}^I_{k,n} \|^2 \le P_k^{\max}, \ \forall k \in\mathcal{K},   \label{eq:Cmm5}\\
        &g(\mathbf{x})\triangleq \ds\eta\sum_{k\in\mathcal{K}} \sum_{n_1\in\mathcal{N}_{1,k}} \| \mathbf{x}^E_{k,n_1} \|^2 +
        (1-\eta)\sum_{k\in\mathcal{K}} \sum_{n\in\mathcal{N}_k} \| \mathbf{x}^I_{k,n} \|^2\le P^{\max} ,  \label{eq:Cmm4} \\
        & p_{k,n_1}(\mathbf{x}^E)- \frac{e_{k,n_1}^{\min}}{\zeta_{k,n_1} \eta} \ge 0, \ \forall k\in\mathcal{K}, n_1\in\mathcal{N}_{1,k}, \label{eq:Cmm6} \\
       &  {\color{black} \| \mathbf{x}^E_{k,n_1} \|^2\leq P_{k}^{\max}, \quad \| \mathbf{x}^I_{k,n} \|^2 \le P_k^{\max},
\ \forall k\in\mathcal{K}, n\in\mathcal{N}_k,}  \label{eq:imp0}
\end{align}
\end{subequations}
where $\mathbf{x}_k\triangleq [\mathbf{x}^E_{k,n_1}; \mathbf{x}^I_{k,n}]_{n_1\in {\cal N}_{1,k}, n\in{\cal N}_k, \in {\cal K}}$.

Constraint \eqref{eq:Cmm5} is the individual cell transmit power budget, $P_k^{\max}$, at each BS $k$
while constraint \eqref{eq:Cmm4} is  the total transmit power budget, $P^{\max}$, of the network. Constraint \eqref{eq:Cmm6} requires that UE $(k,n_1)$ harvests energy is greater than some preset target threshold $e_{k,n_1}^{\min}$. {\color{black}Constraint \eqref{eq:imp0} is imposed to budget the beamforming power separately for each UE $(k,n)$ during both EH and ID times.} Note that the objective \eqref{eq:Omm1} is highly non-concave while constraints \eqref{eq:Cmm5}-\eqref{eq:Cmm6} are non-convex due to coupling between beamforming vectors $\mathbf{x}$ and time splitting factor $\eta$.

\section{Proposed Path-following Computation} \label{sec:prop_0pt}

In order to solve non-convex problem \eqref{eq:P1}, {\color{black}we make the variable change:}
\begin{align} \label{eq:rho_def}
1-\eta = \frac{1}{\mu},
\end{align}
which implies the following linear constraint
\begin{equation}\label{c1}
\mu>1.
\end{equation}
In what follows, we first transform the original max-min secrecy rate problem \eqref{eq:P1} by using a new variable $\mu$.

\textit{Transformation of Problem \eqref{eq:P1} by using a new variable $\mu$}: Using \eqref{eq:rho_def}, the power constraints \eqref{eq:Cmm5} and \eqref{eq:Cmm4} become the following constraints:
\begin{subequations}\label{c22}
\begin{align}
\bar{g}_k(\mathbf{x}_k,\mu)\triangleq \ds\sum_{n_1\in\mathcal{N}_{1,k}} \| \mathbf{x}^E_{k,n_1} \|^2+
\frac{1}{\mu}\sum_{n\in\mathcal{N}_k} \| \mathbf{x}^I_{k,n} \|^2
-\frac{1}{\mu}\sum_{n_1\in\mathcal{N}_{1,k}} \| \mathbf{x}^E_{k,n_1} \|^2 &\le P_k^{\max}, \ \forall k \in\mathcal{K}   \label{c22a}\\
\bar{g}(\mathbf{x},\mu)\triangleq \ds\sum_{k\in\mathcal{K}} \sum_{n_1\in\mathcal{N}_{1,k}} \| \mathbf{x}^E_{k,n_1} \|^2
+\ds\frac{1}{\mu}\sum_{k\in\mathcal{K}} \sum_{n\in\mathcal{N}_k} \| \mathbf{x}^I_{k,n} \|^2 -\frac{1}{\mu}\sum_{k\in\mathcal{K}} \sum_{n_1\in\mathcal{N}_{1,k}} \| \mathbf{x}^E_{k,n_1} \|^2&\le P^{\max}
  \label{c22b}
\end{align}
\end{subequations}
and applying \eqref{eq:rho_def} in \eqref{eq:Cmm6}, the EH constraint \eqref{eq:Cmm6} in variable $\mu$ will become:
\begin{align}\label{eq:Cmm6_2}
p_{k,n_1}(\mathbf{x}^E)\geq \ds\frac{e^{\min}_{k,n_1}}{\zeta_{k,n_1}}\left(1+\frac{1}{\mu-1}\right)-\sigma_a^2.
\end{align}
Under the variable change \eqref{srate2}, the achievable secrecy rate in new variable $\mu$ is given by
\begin{eqnarray}
\bar{f}_{k,n}(\mathbf{x},\mu) &=&  \frac{1}{\mu} f_{k,n}^1(\mathbf{x}^I) - \bar{f}_{k,n}^2(\mathbf{x},\mu)\label{srate3}
\end{eqnarray}
where
\begin{align}\label{eq:fb2}
\bar{f}_{k,n}^2(\mathbf{x},\mu) \triangleq \ln \left(1+\ds\frac{ \|\boldsymbol{\mathcal{H}}_{k,k}^H \mathbf{x}^I_{k,n} \|^2 +\epsilon_{k,k} \|\mathbf{x}^I_{k,n}\|^2 }{\bar{q}_{k,n}(\mathbf{x},\mu) } \right)
\end{align}
and by using $q_{k,n}(\mathbf{x},\eta)$ in \eqref{eq:qk}, $\bar{q}_{k,n}(\mathbf{x},\mu)$ is defined as follows:
\begin{align} \label{eq:qk_beta}
\bar{q}_{k,n}(\mathbf{x},\mu)& \triangleq \ds (\mu-1) \left( \ds\sum_{\bk\in\mathcal{K}} \sum_{\bn\in\mathcal{N}_{1,k}} \|\boldsymbol{\mathcal{H}}_{\bk,k}^H \mathbf{x}^E_{\bk,\bn} \|^2
{\color{black} - }  \ds\sum_{\bk\in\mathcal{K}} \sum_{\bn\in\mathcal{N}_{1,k}} \epsilon_{\bk,k} \|\mathbf{x}^E_{\bk,\bn} \|^2 \right) \notag \\
 &+ \ds\sum_{\bn\in\mathcal{N}_{k} \setminus \{n\}} \|\boldsymbol{\mathcal{H}}_{k,k}^H \mathbf{x}^I_{k,\bn} \|^2 +\ds\sum_{\bk\in\mathcal{K}\setminus\{k\}} \sum_{\bn\in\mathcal{N}_{\bk}} \|\boldsymbol{\mathcal{H}}_{\bk,k}^H \mathbf{x}^I_{\bk,\bn} \|^2 \notag \\
    & {\color{black} - } \left(\ds\sum_{\bn\in\mathcal{N}_{k} \setminus \{n\}} \epsilon_{k,k} \|\mathbf{x}^I_{k,\bn} \|^2+ \ds\sum_{\bk\in\mathcal{K}\setminus\{k\}} \sum_{\bn\in\mathcal{N}_{\bk}}  \epsilon_{\bk,k} \|\mathbf{x}^I_{\bk,\bn} \|^2 \right)  + \mu N_\text{ev}\siga,
 \end{align}

Using \eqref{c22}, \eqref{eq:Cmm6_2}, and \eqref{eq:Cmm6_app}, the equivalence of problem \eqref{eq:P1} in variables $\mathbf{x}$ and $\mu$ is given by
\begin{subequations} \label{eq:P12}
\begin{align}
  \ds  \max_{\mathbf{x}^E_{k,n_1},\mathbf{x}^I_{k,n} \in\mathbb{C}^{M\times 1},\mu} \ &
  \ds\min_{k\in\mathcal{K}, n\in\mathcal{N}_k} \left[ \ds \frac{1}{\mu} f_{k,n}^1(\mathbf{x}^I) - \bar{f}_{k,n}^2(\mathbf{x},\mu) \right]\quad
   \\
  & \text{s.t.}    \quad \quad \quad  {\color{black}  \eqref{eq:imp0} }, \eqref{c1}, \eqref{c22},  \eqref{eq:Cmm6_2}.
\end{align}
\end{subequations}

\textit{Inner Approximation of Power constraint \eqref{c22} and EH constraint \eqref{eq:Cmm6_2}}:\footnote{\color{black}a constraint is called an inner approximation of another constraint if and only if any feasible point
of the former is also feasible for the latter \cite{Tuybook16}}
 Let
$(\mathbf{x}^{(\ell)}, \mu^{(\ell)})$ be a feasible point for (\ref{eq:P12}).
By exploiting the convexity of function $\frac{1}{\mu}\|\mathbf{x}\|^2$, the following inequality
holds true
\begin{equation}\label{aa1}
\frac{\| \mathbf{x}\|^2}{\mu}\geq \frac{2\Re\left\{(\mathbf{x}^{(\ell)})^H\mathbf{x}\right\}}{\mu^{(\ell)}}
-\frac{\| \mathbf{x}^{(\ell)}\|^2}{(\mu^{(\ell)})^2} \mu, \quad \forall \mathbf{x}\in\mathbb{C}^N,
\mathbf{x}^{(\ell)}\in\mathbb{C}^N, \mu>0, \mu^{(\ell)}>0.
\end{equation}
Thus using \eqref{aa1}, an inner convex approximation of non-convex constraints \eqref{c22a} and \eqref{c22b} is given by
\begin{subequations}\label{c22_app}
\begin{eqnarray}
\bar{g}_k^{(\ell)}(\mathbf{x}_k,\mu)\triangleq \ds\sum_{n_1\in\mathcal{N}_{1,k}} \| \mathbf{x}^E_{k,n_1} \|^2+
\frac{1}{\mu}\sum_{n\in\mathcal{N}_k} \| \mathbf{x}^I_{k,n} \|^2 &&\nonumber\\
\ds-\frac{1}{\mu^{(\ell)}}\sum_{n_1\in\mathcal{N}_{1,k}} 2\Re\left\{(\mathbf{x}^{E,(\ell)}_{k,n_1})^H\mathbf{x}^E_{k,n_1}\right\}+\frac{\mu}{(\mu^{(\ell)})^2}\ds\sum_{n_1\in\mathcal{N}_{1,k}} \| \mathbf{x}^{E,(\ell)}_{k,n_1} \|^2
&\le& P_k^{\max}, \ \forall k \in\mathcal{K},   \label{c22a_app}\\
\bar{g}^{(\ell)}(\mathbf{x},\mu)\triangleq \ds\sum_{k\in\mathcal{K}} \sum_{n_1\in\mathcal{N}_{1,k}} \| \mathbf{x}^E_{k,n_1} \|^2
+\ds\frac{1}{\mu}\sum_{k\in\mathcal{K}} \sum_{n\in\mathcal{N}_k} \| \mathbf{x}^I_{k,n} \|^2
&&\nonumber\\
\ds-\frac{1}{\mu^{(\ell)}}\sum_{k\in\mathcal{K}} \sum_{n_1\in\mathcal{N}_{1,k}}2\Re\left\{(\mathbf{x}^{E,(\ell)}_{k,n_1})^H \mathbf{x}^E_{k,n_1}\right\}+\frac{\mu}{(\mu^{(\ell)})^2}\ds\sum_{k\in\mathcal{K}} \sum_{n_1\in\mathcal{N}_{1,k}} \| \mathbf{x}^{E,(\ell)}_{k,n_1} \|^2&\le &P^{\max}. \label{c22b_app}
\end{eqnarray}
\end{subequations}

Next, following the definition of $p_{k,n_1}(\mathbf{x}^E)$ in \eqref{pkn1}, and using the approximation
\begin{equation}\label{boundk1}
| \mathbf{h}_{\bk,k,n}^H \mathbf{x}_{\bk,\bn}  |^2  \geq  -| \mathbf{h}_{\bk,k,n}^H \mathbf{x}_{\bk,\bn}^{(\ell)} |^2 + 2 \Re \left\{ \left(\mathbf{x}_{\bk,\bn}^{(\ell)}\right)^H \mathbf{h}_{\bk,k,n} \mathbf{h}_{\bk,k,n}^H  \mathbf{x}_{\bk,\bn} \right\}, \ \forall \mathbf{x}_{\bar{k},\bar{n}},  \mathbf{x}^{(\ell)}_{\bar{k},\bar{n}}
\end{equation}
an inner approximation of nonconvex constraint \eqref{eq:Cmm6_2} is given by
\begin{align}\label{eq:Cmm6_app}
\ds\sum_{\bk\in\mathcal{K}} \sum_{\bn\in\mathcal{N}_{1,\bk}}\left[2\Re\left\{ \mathbf{h}_{\bk,k,n_1}^H \mathbf{x}^{E,(\ell)}_{\bk,\bn}
\mathbf{h}_{\bk,k,n_1}^H \mathbf{x}^E_{\bk,\bn}\right\} - \left|\mathbf{h}_{\bk,k,n_1}^H \mathbf{x}^{E,(\ell)}_{\bk,\bn}\right|^2\right]
\geq \ds\frac{e^{\min}_{k,n_1}}{\zeta_{k,n_1}}\left(1+\frac{1}{\mu-1}\right)-\sigma_a^2.
\end{align}

Using the convex approximations \eqref{c22_app} and \eqref{eq:Cmm6_app} for the constraints of problem \eqref{eq:P12}, we
obtain {\color{black}the} following inner approximation at $\ell$th iteration:
\begin{subequations} \label{eq:P2}
\begin{align}
  \ds  \max_{\mathbf{x}^E_{k,n_1},\mathbf{x}^I_{k,n} \in\mathbb{C}^{M\times 1},\mu} \ &
  \ds\min_{k\in\mathcal{K}, n\in\mathcal{N}_k} \left[ \ds \frac{1}{\mu} f_{k,n}^1(\mathbf{x}^I) - \bar{f}_{k,n}^2(\mathbf{x},\mu) \right]\quad
   \label{eq:Omm2}\\
  & \text{s.t.}    \quad \quad \quad  {\color{black}  \eqref{eq:imp0} }, \eqref{c1}, \eqref{c22a_app}, \eqref{c22b_app},  \eqref{eq:Cmm6_app}.
\end{align}
\end{subequations}
As observed in \cite{WLP06}, for $\bar{\mathbf{x}}^I_{k,n}=e^{-\jmath.{\sf arg}(\mathbf{h}_{k,k,n}^H
\mathbf{x}^I_{k,n})}\mathbf{x}^I_{k,n}$, one has $|\mathbf{h}_{k,k,n}^H
\mathbf{x}^I_{k,n}|=\mathbf{h}_{k,k,n}^H\bar{\mathbf{x}}^I_{k,n}=\Re\{\mathbf{h}_{k,k,n}^H
\bar{\mathbf{x}}^I_{k,n}\}\geq 0$ and $|\mathbf{h}_{k',k,n'}^H
\mathbf{x}^I_{k,n}|=|\mathbf{h}_{k',k,n'}^H\bar{\mathbf{x}}^I_{k,n}|$ for $(k',n')\neq (k,n)$ and $\jmath \triangleq \sqrt{-1}$.
The problem \eqref{eq:P2} is thus equivalent to the following optimization problem:
\begin{subequations} \label{eq:P3}
\begin{align}
  \ds  \max_{\mathbf{x}^E_{k,n_1},\mathbf{x}^I_{k,n} \in\mathbb{C}^{M\times 1},\mu} & \bar{F}(\mathbf{x},\mu) \triangleq
  \ds\min_{k\in\mathcal{K}, n\in\mathcal{N}_k} \bar{f}_{k,n}(\mathbf{x},\mu) =
  \ds\min_{k\in\mathcal{K}, n\in\mathcal{N}_k} \left[ \ds  \bar{f}_{k,n}^1(\mathbf{x}^I,\mu) - \bar{f}_{k,n}^2(\mathbf{x},\mu) \right]\quad
   \label{eq:Omm3}\\
   & \text{s.t.}  \quad  \Re\left\{\mathbf{h}_{k,k,n}^H \mathbf{x}^I_{k,n}\right\}\geq 0, \ \forall k\in\clK, n\in\clN_k,  \label{eq:Cmm7}\\
&  \quad  {\color{black}  \eqref{eq:imp0} }, \eqref{c22a_app}, \eqref{c22b_app},  \eqref{eq:Cmm6_app}, \eqref{c1},
\end{align}
\end{subequations}
where \vspace{-0.5cm}
\begin{align}\label{eq:fb1}
 \bar{f}_{k,n}^1(\mathbf{x}^I,\mu) \triangleq  \frac{1}{\mu}  \ln \left(1+\ds\frac{(\Re\{\mathbf{h}_{k,k,n}^H \mathbf{x}^I_{k,n} \})^2 - \epsilon_{k,k,n} \| \mathbf{x}_{k,n}^I  \|^2 } {\varphi_{k,n}(\mathbf{x}^I)} \right),
\end{align}

\textit{Lower Approximation of the Objective \eqref{eq:Omm3}}:
{\color{black}For concave lower approximation of $ \bar{f}_{k,n}(\mathbf{x},\mu)$, {\color{black}which agrees with $\bar{f}_{k,n}$
at} $\left(w^{(\ell)}, \mu^{(\ell)}\right)$, we provide a lower bounding concave function
for the first term $\bar{f}_{k,n}^1(\mathbf{x}^I,\mu)$ and an upper bounding convex function for the second term
$\bar{f}_{k,n}^2(\mathbf{x},\mu)$.} Recalling the definition (\ref{eq:varphi}) of
 $\varphi_{k,n}(\mathbf{x}^I)$, we have the following result.
\begin{theorem}
 A lower bounding concave function $\bar{f}_{k,n}^{1,(\ell)}(\mathbf{x}^I,\mu)$ of $\bar{f}_{k,n}^{1}(\mathbf{x}^I,\mu) $,
{\color{black} which agrees with $\bar{f}_{k,n}^1$ at $(\mathbf{x}_{k,n}^{I,(\ell)},\mu^{(\ell)})$},   is given by
\begin{align}\label{eq:f1_kappa}
\bar{f}_{k,n}^{1}(\mathbf{x}^I,\mu) \ge \bar{f}_{k,n}^{1,(\ell)}(\mathbf{x}^I,\mu) \triangleq a^{(\ell)}-b^{(\ell)}\ds\frac{\varphi_{k,n}(\mathbf{x}^I)}{\nu_{k,n}(\mathbf{x}^I_{k,n}) }-c^{(\ell)}\mu
\end{align}
for
\begin{subequations}\label{c7}
\begin{equation}
\nu_{k,n}(\mathbf{x}^I_{k,n})  \le  \psi_{k,n}(\mathbf{x}^I_{k,n}) -  \epsilon_{k,k,n} \| \mathbf{x}_{k,n}^I  \|^2, \ \forall k\in\clK, n\in\clN_k,
\end{equation}
\begin{equation}
\nu_{k,n} \geq 0 \ , \ \ \psi_{k,n} \geq 0 \ ,   \ \forall k\in\clK, n\in\clN_k,
\end{equation}
\end{subequations}
where
\begin{equation}\label{c6}
\begin{array}{c}
a^{(\ell)}=2\ds\frac{\ln(1+d^{(\ell)})}{\mu^{(\ell)}}+\frac{d^{(\ell)}}{\mu^{(\ell)}(d^{(\ell)}+1)}>0, \\
b^{(\ell)}=\ds\frac{(d^{(\ell)})^2}{\mu^{(\ell)}(d^{(\ell)}+1)}>0, \
c^{(\ell)}=\ds\frac{\ln(1+d^{(\ell)})}{(\mu^{(\ell)})^2}>0,\\
d^{(\ell)}=((\Re\{\mathbf{h}_{k,k,n}^H \mathbf{x}^{I,(\ell)}_{k,n}\})^2-\epsilon_0||\mathbf{x}_{k,n}^{I,(\ell)}||^2)
/\varphi_{k,n}(\mathbf{x}^{I,(\ell)})
\end{array}
\end{equation}
and \vspace{-0.5cm}
\begin{equation}\label{ap1}
 \psi_{k,n}(\mathbf{x}^I_{k,n})\triangleq
 2\Re\{\mathbf{h}_{k,k,n}^H \mathbf{x}^{I,(\ell)}_{k,n}\} \Re\left\{\mathbf{h}_{k,k,n}^H \mathbf{x}^I_{k,n}\right\} -
\left(\Re\left\{\mathbf{h}_{k,k,n}^H \mathbf{x}^{I,(\ell)}_{k,n}\right\}\right)^2.
\end{equation}
The upper bounding convex function $\bar{f}_{k,n}^{2,(\ell)}(\mathbf{x},\mu)$ on $\bar{f}_{k,n}^{2}(\mathbf{x},\mu)$,
{\color{black}which agrees with $\bar{f}_{k,n}^{2}$ at $(\mathbf{x}^{(\ell)},\mu^{(\ell})$, } is given by
\begin{align}\label{eq:f2_kappa}
\bar{f}_{k,n}^2(\mathbf{x},\mu)&\leq \bar{f}_{k,n}^{2,(\ell)}(\mathbf{x},\mu) \nonumber \\
&\triangleq \bar{f}_{k,n}^2(w^{(\ell)},\mu^{(\ell)})+ \left(1+\ds\frac{ \|\boldsymbol{\mathcal{H}}_{k,k}^H w_{k,n}^{I,(\ell)} \|^2 +\epsilon_{k,k} \|w_{k,n}^{I,(\ell)} \|^2 }{\bar{q}_{k,n}(w^{(\ell)},\mu^{(\ell)}) }\right)^{-1} \notag \\ & \times
 \left(\ds\frac{ \|\boldsymbol{\mathcal{H}}_{k,k}^H \mathbf{x}_{k,n}^I \|^2 + \epsilon_{k,k} \| \mathbf{x}_{k,n}^{I} \|^2 }{{\color{black}\sqrt{\beta_{k,n}} }}-\ds\frac{ \|\boldsymbol{\mathcal{H}}_{k,k}^H w_{k,n}^{I,(\ell)} \|^2 + \epsilon_{k,k} \|w_{k,n}^{I,(\ell)} \|^2 }{\bar{q}_{k,n}(w^{(\ell)},\mu^{(\ell)}) }\right)
\end{align}
where
\begin{align}
\beta_{k,n} &>0, \ \forall \ k \in \mathcal{K}, \ n \in \mathcal{N}_k  \label{etakn} \\
{\color{black}\sqrt{\beta_{k,n}}} &\leq \bar{q}_{k,n}(\mathbf{x},\mu), \ \forall \ k \in \mathcal{K}, \ n \in \mathcal{N}_k, \label{etakn1}
\end{align}
where constraint \eqref{etakn1}  is {\color{black}innerly approximated by the constraint:}
\begin{eqnarray}
{\color{black}
\ds\frac{1}{2} \left( \frac{\beta_{k,n}}{\sqrt{\beta_{k,n}^{(\ell)}}(\mu^{(\ell)}-1)}
+\frac{\sqrt{\beta_{k,n}^{(\ell)}}(\mu^{(\ell)}-1)}{(\mu-1)^2} \right) \leq \bar{q}_{k,n}^{(\ell)}(\mathbf{x},
\mu)}\label{etakn4}
\end{eqnarray}
and
\begin{equation}\label{etakn5}
2\mu^{(\ell)}-1-\mu>0.
\end{equation}
for
\begin{equation}\label{etakn4a}
{\color{black}\sqrt{\beta_{k,n}^{(\ell)}}=\bar{q}_{k,n}(\mathbf{x}^{(\ell)},\mu^{(\ell)})}
\end{equation}
and
\begin{align}
\bar{q}_{k,n}^{(\ell)}(\mathbf{x},
\mu)&\triangleq \ds
-\frac{1}{\mu-1}\left(\ds\sum_{\bn\in\mathcal{N}_{k} \setminus \{n\}} \epsilon_{k,k} \|\mathbf{x}^I_{k,\bn} \|^2
{\color{black}+} \ds\sum_{\bk\in\mathcal{K}\setminus\{k\}} \sum_{\bn\in\mathcal{N}_{\bk}} \epsilon_{\bk,k} \|\mathbf{x}^I_{\bk,\bn} \|^2 \right)
- \ds\sum_{\bk\in\mathcal{K}} \sum_{\bn\in\mathcal{N}_{1,k}} \epsilon_{\bk,k}  \|\mathbf{x}^E_{\bk,\bn} \|^2\nonumber  \\
&+\ds \sum_{\bk\in\mathcal{K}} \sum_{\bn\in\mathcal{N}_{1,k}}\Re \left\{ \left\la \boldsymbol{\mathcal{H}}_{\bk,k}^H\mathbf{x}^{E,(\ell)}_{\bk,\bn},
2\boldsymbol{\mathcal{H}}_{\bk,k}^H\mathbf{x}^{E,(\ell)}_{\bk,\bn} \mathbf{x}^E_{\bk,\bn}-\boldsymbol{\mathcal{H}}_{\bk,k}^H\mathbf{x}^{E,(\ell)}_{\bk,\bn}  \right\ra \right\}\nonumber\\
&+\ds\frac{2}{\mu^{(\ell)}-1}\left(\ds\sum_{\bn\in\mathcal{N}_{k} \setminus \{n\}} \Re \left\{ \left\la \boldsymbol{\mathcal{H}}_{k,k}^H \mathbf{x}^{I,(\ell)}_{k,\bn}, \boldsymbol{\mathcal{H}}_{k,k}^H \mathbf{x}^I_{k,\bn} \right\ra \right\}\right.\nonumber\\ 
\displaybreak
&\left. +\ds\sum_{\bk\in\mathcal{K}\setminus\{k\}} \sum_{\bn\in\mathcal{N}_{\bk}}\Re \left\{ \left\la \boldsymbol{\mathcal{H}}_{\bk,k}^H \mathbf{x}^{I,(\ell)}_{\bk,\bn} ,\boldsymbol{\mathcal{H}}_{\bk,k}^H \mathbf{x}^I_{\bk,\bn} \right\ra \right\}\right) \nonumber\\
&-\left(\ds\sum_{\bn\in\mathcal{N}_{k} \setminus \{n\}} \|\boldsymbol{\mathcal{H}}_{k,k}^H \mathbf{x}^{I,(\ell)}_{k,\bn} \|^2 +\ds\sum_{\bk\in\mathcal{K}\setminus\{k\}} \sum_{\bn\in\mathcal{N}_{\bk}} \|\boldsymbol{\mathcal{H}}_{\bk,k}^H \mathbf{x}^{I,(\ell)}_{\bk,\bn} \|^2\right)\frac{\mu-1}{(\mu^{(\ell)}-1)^2}\nonumber\\
&+\left(1+\frac{2}{\mu^{(\ell)}-1}-\frac{\mu-1}{(\mu^{(\ell)}-1)^2} \right) N_\text{ev}\siga.\label{etaknb4}
\end{align}
\end{theorem}
\begin{IEEEproof}
See Appendix \ref{app:A}.
\end{IEEEproof}

Thus, by applying Theorem 1, we can use the following convex quadratic program (QP) to achieve minorant maximization for the nonconvex problem \eqref{eq:P3} at feasible $(\mathbf{x}_{k,n}^{E,(\ell)}, \mathbf{x}_{k,n}^{I,(\ell)}, \mu^{(\ell)})$:
 \begin{subequations} \label{eq:P4}
\begin{align}
  \ds  \max_{\mathbf{x}^E_{k,n_1},\mathbf{x}^I_{k,n} \in\mathbb{C}^{M\times 1},\mu} &
  \ds\min_{k\in\mathcal{K}, n\in\mathcal{N}_k} \left[ \ds  \bar{f}_{k,n}^{1,(\ell)}(\mathbf{x}^I,\mu) - \bar{f}_{k,n}^{2,(\ell)}(\mathbf{x},\mu) \right]\quad
   \label{eq:Omm4}\\
    \text{s.t.} \quad & {\color{black}  \eqref{eq:imp0} }, \eqref{c22a_app}, \eqref{c22b_app},  \eqref{eq:Cmm6_app}, \eqref{c1}, \eqref{eq:Cmm7}, \eqref{c7}, \eqref{etakn}, \eqref{etakn4},\eqref{etakn5}.
\end{align}
\end{subequations}

\begin{algorithm}[!t]\caption{Path-following Algorithm for Secrecy Rate Optimization \eqref{eq:P1}}\label{alg:1}
  \begin{algorithmic}[1]
  \State Initialize $\ell := 0$.
  \State Find a feasible point $\left(\mathbf{x}^{E,(0)}, \mathbf{x}^{I,(0)}, \mu^{(0)}\right)$ of \eqref{eq:P12}.
  \Repeat
  \State Solve convex problem \eqref{eq:P4} to find $\left(\mathbf{x}^{E,(\ell+1)}, \mathbf{x}^{I,(\ell+1)}, \mu^{(\ell+1)}\right)$.
  \State Set $\ell := \ell+1$.
  \Until{the objective in \eqref{eq:P12} converges.}
  \end{algorithmic}
\end{algorithm}

\textit{Details of Proposed Algorithm \ref{alg:1} with its Initialization}: The proposed
computation for the max-min secrecy rate problem \eqref{eq:P12} (and hence \eqref{eq:P1}) is summarized in Algorithm~\ref{alg:1}.
{\color{black}Since the objective function in (\ref{eq:P4}) agrees with that in \eqref{eq:P12} at
$(\mathbf{x}^{(\ell)}, \mu^{(\ell)})$, which is also feasible for  (\ref{eq:P4}), it follows that
\begin{equation}\label{eq:conv_proof}
\begin{array}{lll}
\ds\min_{k\in\mathcal{K}, n\in\mathcal{N}_k} \left[ \ds \frac{1}{\mu^{(\ell+1)}}\bar{f}_{k,n}^{1}(\mathbf{x}^{I,(\ell+1)}) - \bar{f}_{k,n}^{2}(\mathbf{x}^{(\ell+1)},\mu^{(\ell+1)}) \right]&\geq&\\
\ds\min_{k\in\mathcal{K}, n\in\mathcal{N}_k} \left[ \ds  \bar{f}_{k,n}^{1,(\ell)}(\mathbf{x}^{I,(\ell+1)},\mu^{(\ell+1)}) - \bar{f}_{k,n}^{2,(\ell)}(\mathbf{x}^{(\ell+1)},\mu^{(\ell+1)}) \right]&>&\\
\ds\min_{k\in\mathcal{K}, n\in\mathcal{N}_k} \left[ \ds  \bar{f}_{k,n}^{1,(\ell)}(\mathbf{x}^{I,(\ell)},\mu^{(\ell)})
 - \bar{f}_{k,n}^{2,(\ell)}(\mathbf{x}^{(\ell)},\mu^{(\ell)}) \right]&=&\\
 \ds\min_{k\in\mathcal{K}, n\in\mathcal{N}_k} \left[ \ds \frac{1}{\mu^{(\ell)}}\bar{f}_{k,n}^{1}(\mathbf{x}^{I,(\ell)}) - \bar{f}_{k,n}^{2}(\mathbf{x}^{(\ell)},\mu^{(\ell)}) \right],&&
\end{array}
\end{equation}
i.e. $(\mathbf{x}^{(\ell+1)},\mu^{(\ell+1)})$ is a feasible point, which is better than  $(\mathbf{x}^{(\ell)},\mu^{(\ell)})$ for \eqref{eq:P12}, whenever $(\mathbf{x}^{(\ell+1)},\mu^{(\ell+1)})\neq (\mathbf{x}^{(\ell)},\mu^{(\ell)})$. On the other hand, if $(\mathbf{x}^{(\ell+1)},\mu^{(\ell+1)})= (\mathbf{x}^{(\ell)},\mu^{(\ell)})$, i.e. $(\mathbf{x}^{(\ell)},\mu^{(\ell)})$ is the optimal solution of
the convex optimization problem (\ref{eq:P4}) then it must satisfy the
{\color{black}first order necessary optimality condition} for (\ref{eq:P4}), which
obviously is also {\color{black}the first order necessary optimality condition}
 for \eqref{eq:P12}. We thus proved that the sequence
$\{ (\mathbf{x}^{(\ell)},\mu^{(\ell)})\}$ converges to a point satisfying
{\color{black}the first order necessary optimality condition}
for the nonconvex optimization problem \eqref{eq:P12}.}

 A feasible point $\left(\mathbf{x}^{E,(0)}, \mathbf{x}^{I,(0)}, \mu^{(0)}\right)$ for \eqref{eq:P12} (and hence \eqref{eq:P1})
 for initializing Algorithm \ref{alg:1} is found as  as follows.
 We first fix $\mu^{(0)}$ and solve the following convex problem:
\begin{subequations}\label{c8i.al}
\begin{eqnarray}
\ds\max_{\mathbf{x}^x_{k,n}\in\mathbb{C}^{M\times 1}, x\in\{I,E\}}\min_{k\in\clK, n\in\clN_k}
\Re\left\{\mathbf{h}_{k,k,n}^H \mathbf{x}^E_{k,n_1}\right\}
-\sqrt{e^{\min}_{k,n}/\left(\zeta_{k,n}\left(1-1/\mu^{(0)}\right) \right)} \label{c8i.al.ob} \\
\mbox{s.t.}\quad \| \mathbf{x}^E_{k,n_1} \|^2\leq P_{k}^{\max}, \quad \| \mathbf{x}^I_{k,n} \|^2 \le P_k^{\max},
\ \forall k\in\mathcal{K}, n\in\mathcal{N}_k, \label{imp}
\\
\Re\left\{\mathbf{h}_{k,k,n}^H {\color{black}\mathbf{x}^I_{k,n}}\right\} \ge \sqrt{e^{r^{\min}\mu^{(0)}}-1 }  \left\| \begin{matrix} \sigma_a\cr
\left(\mathbf{h}_{\bk,k,n}^H {\color{black}\mathbf{x}^I_{\bk,\bn}} \right)_{\bar{k},\bar{n}\in \mathcal{K},\mathcal{N} \setminus \{k,n\} }
\end{matrix}
\right\|_{2},  \ k \in \mathcal{K}, n\in\mathcal{N}, \label{c8ial_c1}\\
\ds \bar{g}_k \left( \mathbf{x}_k,\mu^{(0)} \right) \le P_k^{\max}, \ \forall k \in\mathcal{K}, \label{c8i.al_c2}\\
\ds\bar{g} \left(\mathbf{x},\mu^{(0)} \right) \le P^{\max}, \label{c8i.al_c3}
\end{eqnarray}
\end{subequations}
where, for quick convergence, the constraint \eqref{c8ial_c1} on the information rate of UE $(k,n)$ is imposed.
{\color{black}Note that the constraint (\ref{eq:Cmm6_2})  is satisfied if the objective function (\ref{c8i.al.ob}) is positive.}
The constraint \eqref{c8ial_c1} is a second-order cone constraint \cite{Nasir-16-CL-A}.
Using the optimal solution  $\mathbf{x}^{E,(0)}_{k,n}$  of \eqref{c8i.al} as
the initial point, {\color{black}we then iteratively solve the following convex program:
\begin{align}\label{c8i.al2}
\ds\max_{\mathbf{x}^x_{k,n}\in\mathbb{C}^{M\times 1}, x\in\{I,E\}}\min_{k\in\clK, n\in\clN_k}
\ds\sum_{\bk\in\mathcal{K}} \sum_{\bn\in\mathcal{N}_{\bk}}\left[2\Re\left\{ \mathbf{h}_{\bk,k,n}^H \mathbf{x}^{E,(\ell)}_{\bk,\bn}
\mathbf{h}_{\bk,k,n}^H \mathbf{x}^E_{\bk,\bn}\right\} - \left|\mathbf{h}_{\bk,k,n}^H \mathbf{x}^{E,(\ell)}_{\bk,\bn}\right|^2\right]
\notag \\ - \ds\frac{e^{\min}_{k,n}}{\zeta_{k,n}}\left(1+\frac{1}{\mu^{(0)}-1}\right)-\sigma_a^2  \quad\mbox{s.t.}\quad \eqref{imp},\eqref{c8ial_c1},\eqref{c8i.al_c2},\eqref{c8i.al_c3}.
\end{align}
until a positive value of the objective function is achieved. If either problem \eqref{c8i.al}
is found infeasible or an positive  value by solving \eqref{c8i.al2} is not found, we use different value of $\mu^{(0)}$ and repeat the above process until a feasible point $\left(\mathbf{x}^{E,(0)}, \mathbf{x}^{I,(0)}, \mu^{(0)}\right)$ is obtained.}\footnote{{\color{black}Our simulation results in Sec.~\ref{sec:sim} show that the initialization problems \eqref{c8i.al} or \eqref{c8i.al2} are feasible, and in almost all of the scenarios considered, we achieve a positive optimal value of \eqref{c8i.al2} in one single iteration and with the first tried value of $\mu^{(0)} = 1.11$.}}

\section{Energy Efficient Secure Beamforming} \label{sec:EE}
This section extends the proposed robust secrecy rate maximization algorithm to solve the secrecy energy efficiency (SEE) maximization problem, which is formulated  in the presence of channel estimation errors and eavesdroppers as
\begin{subequations} \label{eP1}
\begin{align}
  \ds  \max_{\overset{\mathbf{x}^E_{k,n_1},\mathbf{x}^I_{k,n} \in\mathbb{C}^{M\times 1}}{\eta\in(0,1)}} &
  \ds\min_{k\in\mathcal{K}}\frac{\ds\sum_{n\in\mathcal{N}_k}[ (1-\eta)f_{k,n}^1(\mathbf{x}^I) -
  f_{k,n}^2(\mathbf{x},\eta)]}{ \frac{1}{\xi}g_k(\mathbf{x}_k,\eta)+MP_A+P_c  }\quad\text{s.t.} \quad
    (\ref{eq:Cmm5}), (\ref{eq:Cmm4}), (\ref{eq:Cmm6}), {\color{black}(\ref{eq:imp0}  )} \label{eP1a}\\
    \quad & \ds (1-\eta)f_{k,n}^1(\mathbf{x}^I) -
  f_{k,n}^2(\mathbf{x},\eta)\geq r_{k,n},\quad \forall\ k\in\mathcal{K}, n\in\mathcal{N}_k,\label{eP1b}
\end{align}
\end{subequations}
where $\xi$ is the constant power amplifier efficiency, $P_A$ is the power dissipation at each transmit antenna, $P_c$ is the fixed circuit power consumption for base-band processing and $r_{k,n}$ is the threshold secrecy rate to ensure quality of service. The security and energy efficiency
are combined  into a single objective in (\ref{eP1a}) to express  the so-called secrecy EE (SEE)
in terms of secrecy bits per Joule per Hertz.

The conventional approach to address (\ref{eP1})
(see e.g. \cite{ZJ15,Zetal16}) is based on the Dinkelbach's method of fractional programming \cite{D67} to find
$\tau>0$ such that the optimal value of the following optimization problem is zero
\begin{eqnarray}\label{d1}
 \ds  \max_{\overset{\mathbf{x}^E_{k,n_1},\mathbf{x}^I_{k,n} \in\mathbb{C}^{M\times 1}}{\eta\in(0,1)}}\
\ds\min_{k\in\mathcal{K}}\left\{ \ds\sum_{n\in\mathcal{N}_k}[ (1-\eta)f_{k,n}^1(\mathbf{x}^I) -
  f_{k,n}^2(\mathbf{x},\eta)]-\tau[\frac{1}{\xi}g_k(\mathbf{x}_k,\eta)+MP_A+P_c]\right\}\nonumber\\
  \quad\text{s.t.} \quad
    (\ref{eq:Cmm5}), (\ref{eq:Cmm4}), (\ref{eq:Cmm6}),  {\color{black}(\ref{eq:imp0}  )}, (\ref{eP1b}).
\end{eqnarray}
However, for each fixed $\tau>0$, the optimization problem (\ref{d1}) is highly nonconvex convex  and thus is still difficult computationally. It is important to realize that the original Dinkelbach's method \cite{D67} is attractive only
for maximizing a ratio of a convex and concave functions over a convex set, under which
each subproblem for fixed $\tau$ is an easy convex optimization problem. It is hardly useful whenever
either the objective is not ratio of a concave and convex function or the constrained set is
not convex.

We now develop an efficient path-following computational procedure for solution of (\ref{eP1}), which bypasses such difficult optimization
problem (\ref{d1}). Using the variable change \eqref{c1} again, this problem is equivalent to
\begin{subequations} \label{eP2}
\begin{align}
  \ds  \max_{\overset{\mathbf{x}^E_{k,n_1},\mathbf{x}^I_{k,n} \in\mathbb{C}^{M\times 1}}{\mu>1, t_k>0}} &
  \ds\min_{k\in\mathcal{K}}\ \ds\sum_{n\in\mathcal{N}_k}[\frac{f_{k,n}^1(\mathbf{x}^I)}
  {\mu\sqrt{t_k}} -
  \frac{\bar{f}_{k,n}^2(\mathbf{x},\mu)}{\sqrt{t_{k}}}]\quad\text{s.t.} \quad {\color{black}(\ref{eq:imp0}  )},
   \eqref{c22},  \eqref{eq:Cmm6_2}, \eqref{c1},    \label{eP2a}\\
    \quad & \ds \bar{f}_{k,n}^1(\mathbf{x}^I,    \mu) -
  \bar{f}_{k,n}^2(\mathbf{x},\mu)\geq r_{k,n}\quad \forall\ k\in\mathcal{K}, n\in\mathcal{N}_k,\label{eP2b}\\
  & \frac{1}{\xi}\bar{g}_k(\mathbf{x}_k,\mu)+MP_A+P_c\leq \sqrt{t_{k}}, \forall\ k\in {\cal K},\label{eP2c}
\end{align}
\end{subequations}
By using (\ref{eq:f1_kappa}) we obtain
\begin{eqnarray}
\ds\frac{f_{k,n}^1(\mathbf{x}^I)}{\mu\sqrt{t_k}}&\geq&
A^{(\ell)}-B^{(\ell)}\ds\frac{\varphi_{k,n}(\mathbf{x}^I)}{\nu_{k,n}(\mathbf{x}^I_{k,n}) }-C^{(\ell)}\mu
\sqrt{t_k} \notag \\
&\geq&\Phi^{(\ell)}_{k,n}(\mathbf{x},\mu,t_k) \notag \\
&\triangleq&A^{(\ell)}-B^{(\ell)}\ds\frac{\varphi_{k,n}(\mathbf{x}^I)}{\nu_{k,n}(\mathbf{x}^I_{k,n}) }-
C^{(\ell)}\left(\frac{\sqrt{t_k^{(\ell)}}}{2\mu^{(\ell)}}\mu^2+\frac{\mu^{(\ell)}}{2\sqrt{t_k^{(\ell)}}}t_k \right) \label{eq:phiEE}
\end{eqnarray}
for (\ref{c7}), where $\sqrt{t_k^{(\ell)}}=\bar{g}_k(\mathbf{x}^{(\ell)}_k,\mu^{(\ell)})+MP_A+P_c$ and
\[
\begin{array}{c}
A^{(\ell)}=2\ds\frac{\ln(1+D^{(\ell)})}{\mu^{(\ell)}\sqrt{t_k^{(\ell)}}}+\frac{D^{(\ell)}}
{\mu^{(\ell)}\sqrt{t_k^{(\ell)}}(D^{(\ell)}+1)}>0, \\
B^{(\ell)}=\ds\frac{(D^{(\ell)})^2}{\mu^{(\ell)}\sqrt{t_k^{(\ell)}}(D^{(\ell)}+1)}>0, \\
C^{(\ell)}=\ds\frac{\ln(1+D^{(\ell)})}{[\mu^{(\ell)}\sqrt{t_k^{(\ell)}}]^2}>0,\\
D^{(\ell)}=((\Re\{\mathbf{h}_{k,k,n}^H \mathbf{x}^{I,(\ell)}_{k,n}\})^2-\epsilon_0||\mathbf{x}_{k,n}^{I,(\ell)}||^2)
/\varphi_{k,n}(\mathbf{x}^{I,(\ell)}).
\end{array}
\]
Similarly to (\ref{eq:f2_kappa}), we have
\begin{eqnarray}
\ds\frac{\bar{f}_{k,n}^2(\mathbf{x},\mu)}{\sqrt{t_{k}}}&\leq&
\ds\frac{\bar{f}_{k,n}^2(w^{(\ell)},\mu^{(\ell)})}{\sqrt{t_{k}}}+ \left(1+\ds\frac{ \|\boldsymbol{\mathcal{H}}_{k,k}^H w_{k,n}^{I,(\ell)} \|^2 +\epsilon_{k,k} \|w_{k,n}^{I,(\ell)} \|^2 }{\bar{q}_{k,n}(w^{(\ell)},\mu^{(\ell)}) }\right)^{-1} \notag \\
&& \times
 \left(\ds\frac{ \|\boldsymbol{\mathcal{H}}_{k,k}^H \mathbf{x}_{k,n}^I \|^2 + \epsilon_{k,k} \| \mathbf{x}_{k,n}^{I} \|^2 }{\sqrt{t_{k}\beta_{k,n}} }-\ds\frac{ \|\boldsymbol{\mathcal{H}}_{k,k}^H w_{k,n}^{I,(\ell)} \|^2 + \epsilon_{k,k} \|w_{k,n}^{I,(\ell)} \|^2 }{\bar{q}_{k,n}(w^{(\ell)},\mu^{(\ell)}) \sqrt{t_{k}}}\right)\label{ef2}\\
 &\leq&\Psi^{(\ell)}_{k,n}(\mathbf{x},\mu,t_k),\label{ef3}
\end{eqnarray}
with (\ref{etakn}), (\ref{etakn1}) and
\begin{equation}\label{ef4}
0<t_{k}\leq 3t^{(\ell)}_{k},\ \forall\ k\in {\cal K},
\end{equation}
where
\begin{eqnarray}
\Psi^{(\ell)}_{k,n}(\mathbf{x},\mu,t_k)&\triangleq&\ds\frac{\bar{f}_{k,n}^2(w^{(\ell)},\mu^{(\ell)})}{\sqrt{t_{k}}}+
\left(1+\ds\frac{ \|\boldsymbol{\mathcal{H}}_{k,k}^H w_{k,n}^{I,(\ell)} \|^2 +\epsilon_{k,k} \|w_{k,n}^{I,(\ell)} \|^2 }{\bar{q}_{k,n}(w^{(\ell)},\mu^{(\ell)}) }\right)^{-1} \notag \\
&& \hspace{-1.5cm} \times
 \left(\ds\frac{ \|\boldsymbol{\mathcal{H}}_{k,k}^H \mathbf{x}_{k,n}^I \|^2 + \epsilon_{k,k} \| \mathbf{x}_{k,n}^{I} \|^2 }{\sqrt{t_{k}\beta_{k,n}} }-\ds\frac{ \|\boldsymbol{\mathcal{H}}_{k,k}^H w_{k,n}^{I,(\ell)} \|^2 + \epsilon_{k,k} \|w_{k,n}^{I,(\ell)} \|^2 }{2\bar{q}_{k,n}(w^{(\ell)},\mu^{(\ell)})\sqrt{t^{(\ell)}_{k}}}
 \left(3-\frac{t_{k}}{t_{k}^{(\ell)}}  \right)\right)\label{ef5}
\end{eqnarray}
For the approximation \eqref{ef3} under \eqref{ef4}, we have used the following inequality
\[
\frac{1}{\sqrt{t}}\geq \frac{1}{2\sqrt{\bar{t}}} \left(3-\frac{t}{\bar{t}}\right)\ \forall\ t>0, \bar{t}>0.
\]
The inner approximations in \eqref{eq:phiEE} and \eqref{ef3} can be easily followed by using the procedure in Appendix \ref{app:A}. The following convex program is minorant maximization for the nonconvex program (\ref{eP2})
\begin{subequations} \label{eP3}
\begin{align}
  \ds  \max_{\overset{\mathbf{x}^E_{k,n_1},\mathbf{x}^I_{k,n} \in\mathbb{C}^{M\times 1}}{\mu>1,t_k>0}} &
  \ds\min_{k\in\mathcal{K}}\ \ds\sum_{n\in\mathcal{N}_k}[\Phi^{(\ell)}_{k,n}(\mathbf{x},\mu,t_k)) -
  \Psi^{(\ell)}_{k,n}(\mathbf{x},\mu,t_k)] \label{eP3a}\\
  \text{s.t.}\quad&{\color{black}\eqref{eq:imp0}  },
    \eqref{c22a_app}, \eqref{c22b_app},  \eqref{eq:Cmm6_app}, \eqref{c1}, \eqref{eq:Cmm7}, \eqref{c7}, \eqref{etakn}, \eqref{ef4},\eqref{etakn4},\eqref{etakn5}, \label{eP3b}\\
    \quad & \ds \bar{f}_{k,n}^{1,(\ell)}(\mathbf{x}^I,    \mu) -
  \bar{f}_{k,n}^{2,(\ell)}(\mathbf{x},\mu)\geq r_{k,n}\quad \forall\ k\in\mathcal{K}, n\in\mathcal{N}_k,\label{eP3c}\\
  & \frac{1}{\xi}\bar{g}^{(\ell)}_k(\mathbf{x}_k,\mu)+MP_A+P_c\leq \sqrt{t_{k}}, \forall\ k\in {\cal K}.\label{eP3d}
\end{align}
\end{subequations}

\begin{algorithm}[!t]\caption{Path-following Algorithm for SEE Optimization \eqref{eP1}}\label{alg:2}
  \begin{algorithmic}[1]
  \State Initialize $\ell := 0$.
  \State Find a feasible point $\left(\mathbf{x}^{E,(0)}, \mathbf{x}^{I,(0)},\mathbf{t}^{(0)}, \mu^{(0)}\right)$ of
  \eqref{eP2}.
  \Repeat
  \State Solve convex program \eqref{eP3} for $\left(\mathbf{x}^{E,(\ell+1)}, \mathbf{x}^{I,(\ell+1)}, \mathbf{t}^{(\ell+1)} \mu^{(\ell+1)}\right)$.
  \State Set $\ell := \ell+1$.
  \Until{the objective in \eqref{eP2} converges.}
  \end{algorithmic}
\end{algorithm}

Algorithm \ref{alg:2} outlines the steps to solve
max-min energy efficiency problem \eqref{eP2} (and hence \eqref{eP1}). Similar to Algorithm~\ref{alg:1},
{\color{black}Algorithm \ref{alg:2} generates a sequence $\left\{\left(\mathbf{x}^{E,(\ell)}, \mathbf{x}^{I,(\ell)}, \mathbf{t}^{(\ell)}, \mu^{(\ell)}\right)\right\}$
of improved points of \eqref{eP3}, which converges to a KKT point,} where $\mathbf{t}^{(\ell)} \triangleq [ t_1^{(\ell)},\hdots,t_K^{(\ell)} ]^T$.  A feasible point $\left(\mathbf{x}^{E,(0)}, \mathbf{x}^{I,(0)}, \mathbf{t}^{(0)}, \mu^{(0)}\right)$ of \eqref{eq:P2} (and hence \eqref{eP1}) for initializing Algorithm~\ref{alg:2}
can be obtained by first solving \eqref{c8i.al} and \eqref{c8i.al2} followed by a feasibility problem \eqref{eP3b}, \eqref{eP3c}, and \eqref{eP3d}. It was already reported in Section \ref{sec:prop_0pt} that how efficiently the solution of \eqref{c8i.al} and \eqref{c8i.al2} is obtained. The solution to the feasibility problem \eqref{eP3b}, \eqref{eP3c}, and \eqref{eP3d} is mostly obtained at the first iteration.

\section{Simulation Results} \label{sec:sim}

\ifCLASSOPTIONpeerreview
\begin{figure*}[t]
    \centering
    \begin{minipage}[h]{0.48\textwidth}
    \centering
    \includegraphics[width=1.01 \textwidth]{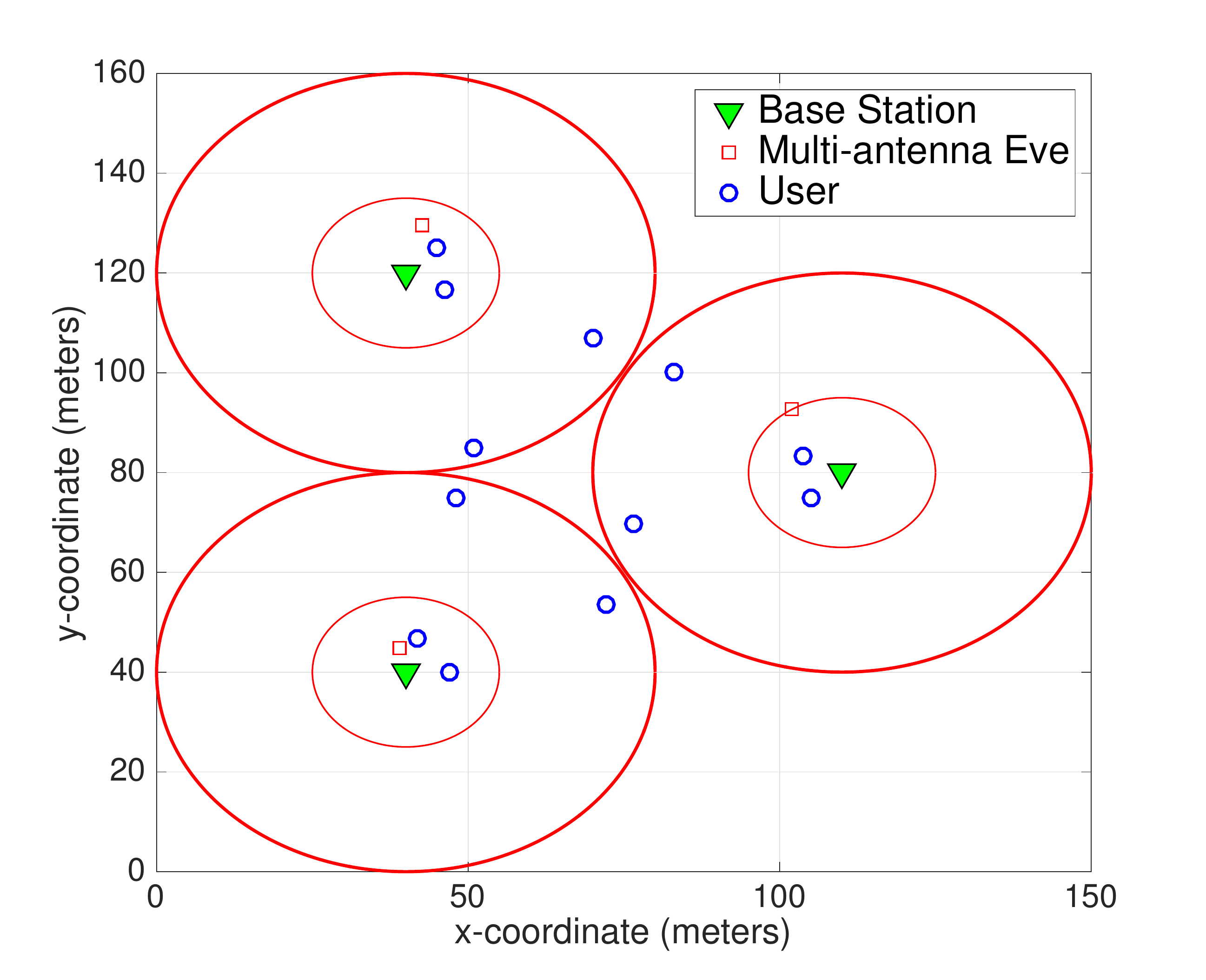}
  \caption{A multicell network setup used in our numerical examples}
  \label{fig:nw_top}
  \end{minipage}
    \hspace{0.3cm}
    \begin{minipage}[h]{0.48\textwidth}
    \centering
    \includegraphics[width=1.01 \textwidth]{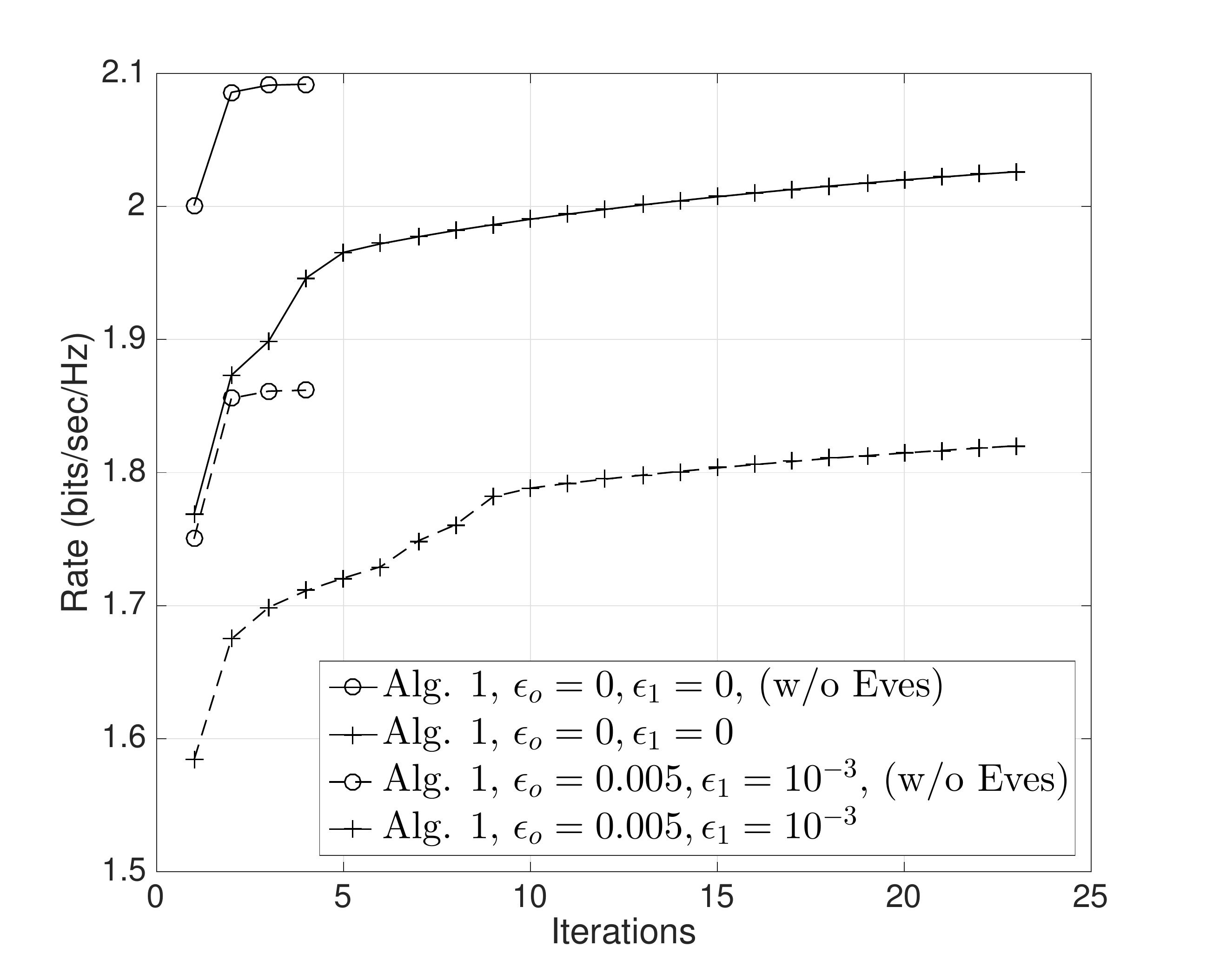}
  \caption{Convergence of Algorithm \ref{alg:1} for $M=5$ and $e^{\min} = -20$ dBm. }
  \label{fig:conv}
  \end{minipage}
\end{figure*}
\else
\fi

To analyze the proposed algorithms through simulations,  a network topology as shown in Fig.~\ref{fig:nw_top} is set up.
There are $K = 3$ cells and $N = N_k = 4, \ \forall \ k \in \mathcal{K}$ UEs per cell with two
 placed inside the inner-circle in zone-1 and the remaining two  placed in the outer zone near cell-edges, i.e., $N_{1,k} = N_{2,k} = 2$, $\forall$ $k$. The cell radius is set to be $40$m with inner circle radius of $15$m. A single $N_\text{ev}=2$-antennas eavesdropper is randomly placed
 inside the inner circle in each cell. The path loss exponent is set to be $\mu = 3$. We generate Rician fading channels with Rician factor, $K_R=10$ dB \cite{Krikidis-14-MCOM-A}.
 For simplicity, {\color{black}set $e_{k,n_1}^{\min} \equiv e^{\min}$ for the energy harvesting thresholds
and $\zeta_{k,n_1} \equiv \zeta$, $\ \forall k,n_1$ for the energy harvesting conversion.} Further, we set energy conversion efficiency $\zeta = 0.5$, noise variance $\siga = -90$ dBm {\color{black}(unless specified otherwise)},  maximum BS transmit power $P_k^{\max} =  26$ dBm, $\forall$ $k$,
 which is in line with the frequently made assumption for the power budget of small-cell BSs \cite{Onireti-15-A}. We choose the value $P^{\max} = 30$ dBm of the power budget for the whole network. Unless stated otherwise, we choose the uncertainty in eavesdroppers' and neighboring users' channels, $\epsilon_0 = 0.005$ and we choose the uncertainty in serving users' channels $\epsilon_1 = 10^{-3}$. It is justified to assume that BSs can achieve good channel estimates for their serving cell users compared to the neighboring cell users in a dense small cell network. Later in this section, we also investigate the effect of different values of channel uncertainties on the achievable secrecy rate. For energy efficiency maximization problem in Section \ref{sec:EE}, we choose power amplifier efficiency $\xi = 0.2$,   is the power dissipation at each transmit antenna $P_A = 0.6$W ($27.78$  dBm), and circuit power consumption $P_c = 2.5$W ($33.97$ dBm)  \cite{Imran-11-EPD-A,Leng-15-ICNC-P}. We set the threshold secrecy rate $r_{k,n} = 0.1$ bits/sec/Hz for $M = 4$ antennas at the BS and otherwise $r_{k,n} = 0.5$ bits/sec/Hz for $M \in \{5,6\}$ antenna-BSs.


The convergence of  Algorithm \ref{alg:1} for $M=5$ antenna BS and minimum energy harvesting threshold $e^{\min} = -20$ dBm
{\color{black} is shown by Fig. \ref{fig:conv}}.
 We can see that for some fixed channel, whether we assume perfect channel estimation $\epsilon_0 = 0$, $\epsilon_1 = 0$ or assume some channel uncertainty $\epsilon_0 = 0.005$, $\epsilon_0 = 10^{-3}$, Algorithm 1 converges within $20-25$ iterations. We also observe that if we assume the absence of eavesdropper,  Algorithm 1 quickly converges in about 4 iterations. {\color{black}On average, Algorithm 1} requires $22.5$ iterations before convergence, while the absence of eavesdroppers drops down the average required number of iterations to $3.5$. The slower convergence in the presence of eavesdroppers is expected since then, not only the objective \eqref{eq:Omm4} gets quite complicated, but also new constraints, \eqref{etakn}, \eqref{etakn4}, and \eqref{etakn5} are required to be satisfied.


\ifCLASSOPTIONpeerreview
\begin{figure*}[t]
    \centering
    \begin{minipage}[h]{0.48\textwidth}
    \centering
    \includegraphics[width=1.01 \textwidth]{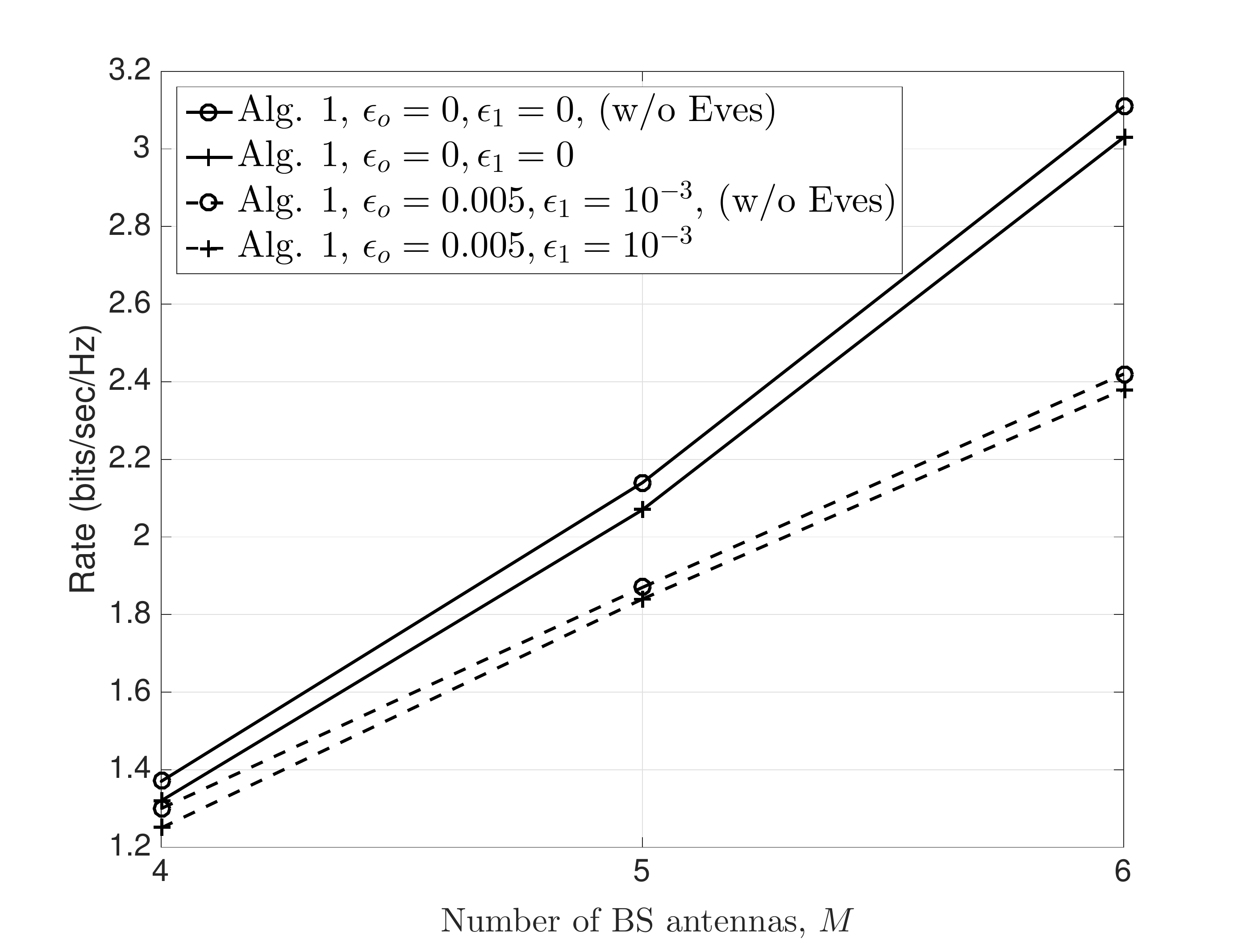}
  \caption{Robust secrecy rate and normal rate in the presence and absence of eavesdroppers, respectively, for varying number of antennas $M$ and different levels of channel uncertainties, but fixed EH threshold $e^\text{min} = -20$ dBm.}
  \label{fig:M}
  \end{minipage}
    \hspace{0.3cm}
    \begin{minipage}[h]{0.48\textwidth}
    \centering
    \includegraphics[width=1.01 \textwidth]{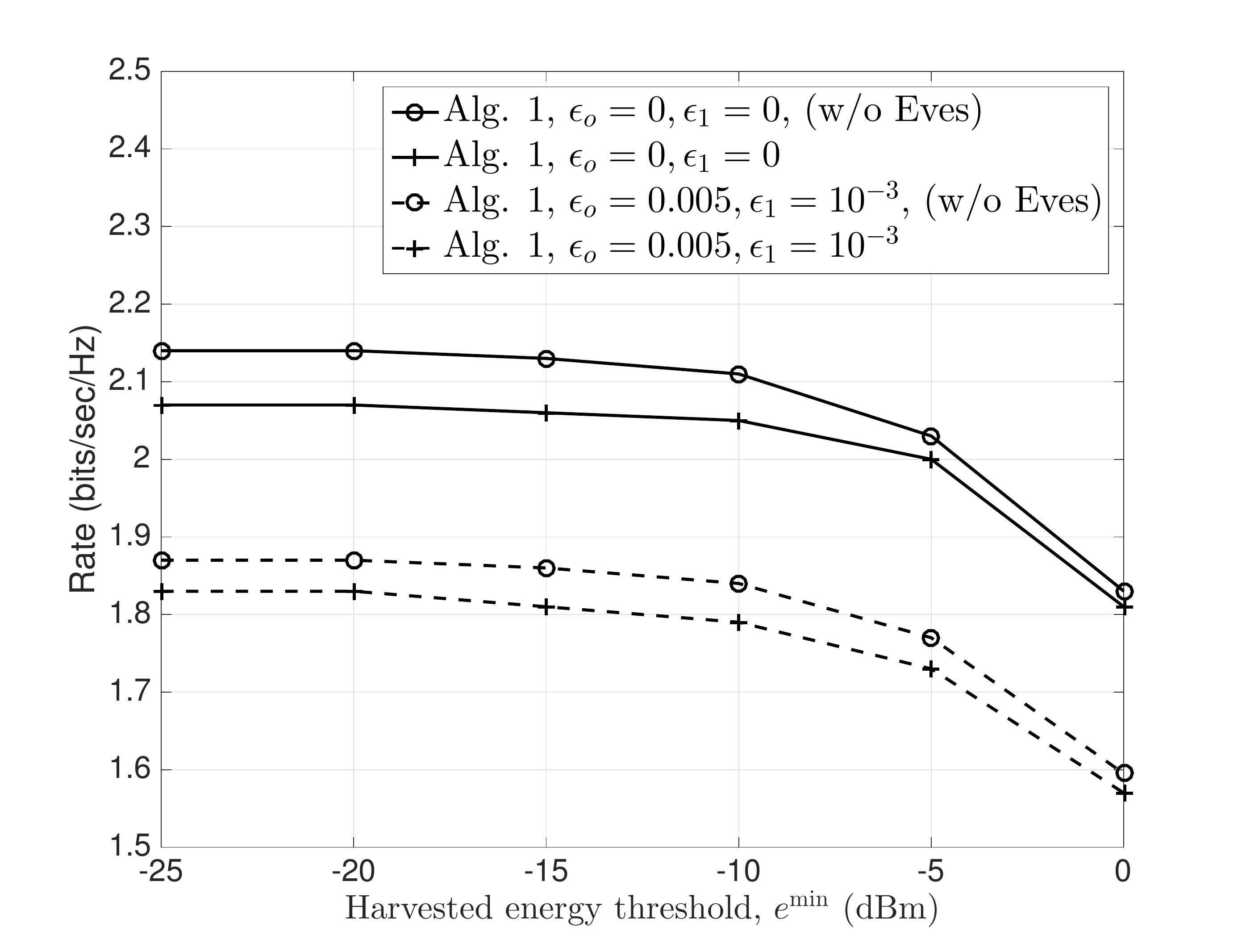}
  \caption{Robust secrecy rate and normal rate in the presence and absence of eavesdroppers, respectively, for different values of EH thresholds $e^\text{min}$ and different levels of channel uncertainties, but fixed number of BS antennas $M=5$.}
  \label{fig:H}
  \end{minipage}
\end{figure*}
\else
\fi

 {\color{black}The worst secrecy  and  normal rates} (in the absence of eavesdroppers) for both perfect channel estimation $\epsilon_0 = 0$, $\epsilon_1 = 0$ and with the presence of channel uncertainty of $\epsilon_0 = 0.005$, $\epsilon_1 \in 10^{-3}$ are provided by Figs. \ref{fig:M} and \ref{fig:H}.
  Basically, the secrecy rate by Algorithm \ref{alg:1}
 is compared with the normal rate. The latter excludes eavesdroppers and
 accordingly the optimization problem \eqref{eq:P1} with $f_{k,n}^2(\mathbf{x},\eta)\equiv 0$ in \eqref{eq:Omm1}
 is solved. The dashed curves in Figs. \ref{fig:M} and \ref{fig:H} have been plotted to refer to the presence of channel uncertainties, while solid line curves refer to the absence of channel uncertainty $\epsilon_0 = 0$, $\epsilon_1 = 0$. We can observe from Figs. \ref{fig:M} and \ref{fig:H} that the proposed robust secrecy rate algorithm in the presence of channel uncertainties perform quite well and close to the case that assumes perfect channel estimation. However, the performance gap increases by increasing the number of antennas as can be seen from Fig. \ref{fig:M}. It is expected because increasing the number of antennas, say from $M=4$ to $M=5$ increases the channel uncertainty in additional $KN = 12$ channel co-efficients. Moreover, we observe that the optimized rate by the proposed Algorithm \ref{alg:1} is quite close to that achieved by the modified algorithm, which assumes absence of eavesdroppers in Algorithm \ref{alg:1}. Fig. \ref{fig:M} plots the rate for different number of BS antennas $M \in \{4,5,6 \}$ with fixed EH threshold $e^{\min} = -20$ dBm, while Fig. \ref{fig:H} plots the rate for varying values of EH targets $e^{\min} \in \{-25,-20,\hdots,0 \}$ dBm with fixed number of antennas at the BS $M=5$. In Fig. \ref{fig:M}, we observe almost linear increase in the achievable rate with increase in the number of BS antennas. In Fig. \ref{fig:H}, we observe decrease in the achievable rate with the increase in the EH targets. This is because higher EH targets demands for more power from the BS to perform energy harvesting, which results in the decrease in the available power for information decoding, thus decreases the achievable information rate. Overall, Figs. \ref{fig:M} and \ref{fig:H} indicate the robustness of our proposed Algorithm \ref{alg:1}.

   \ifCLASSOPTIONpeerreview
\begin{figure*}[t]
    \centering
    \begin{minipage}[h]{0.48\textwidth}
    \centering
    \includegraphics[width=1.01 \textwidth]{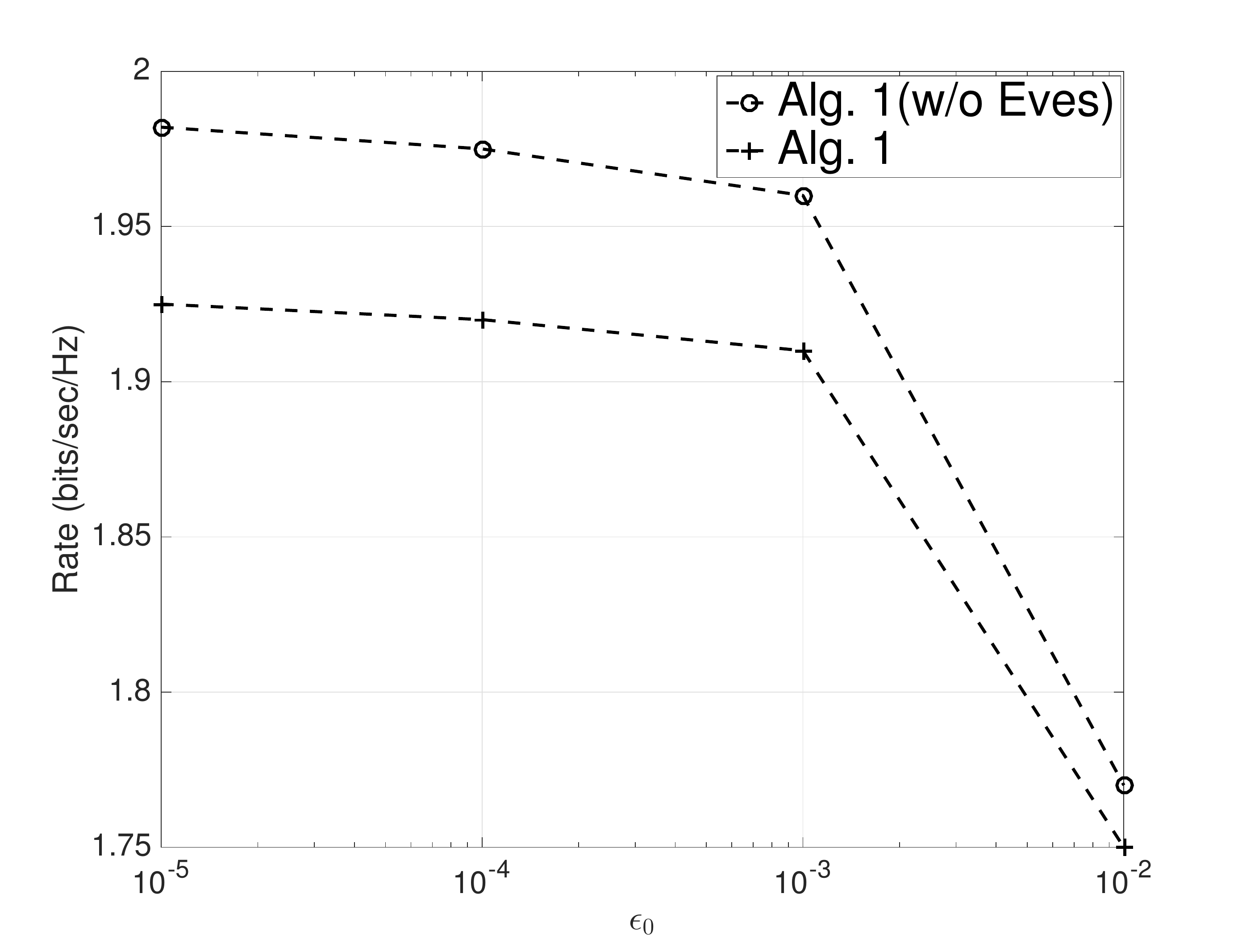}
  \caption{Robust secrecy rate (with eavesdroppers) and normal rate (without eavesdroppers) versus different levels of channel uncertainty $\epsilon_0$ for fixed energy harvesting threshold $e^\text{min} = -20$ dBm and BS antennas $M=5$.}
  \label{fig:Ue0}
  \end{minipage}
    \hspace{0.3cm}
    \begin{minipage}[h]{0.48\textwidth}
    \centering
    \includegraphics[width=1.01 \textwidth]{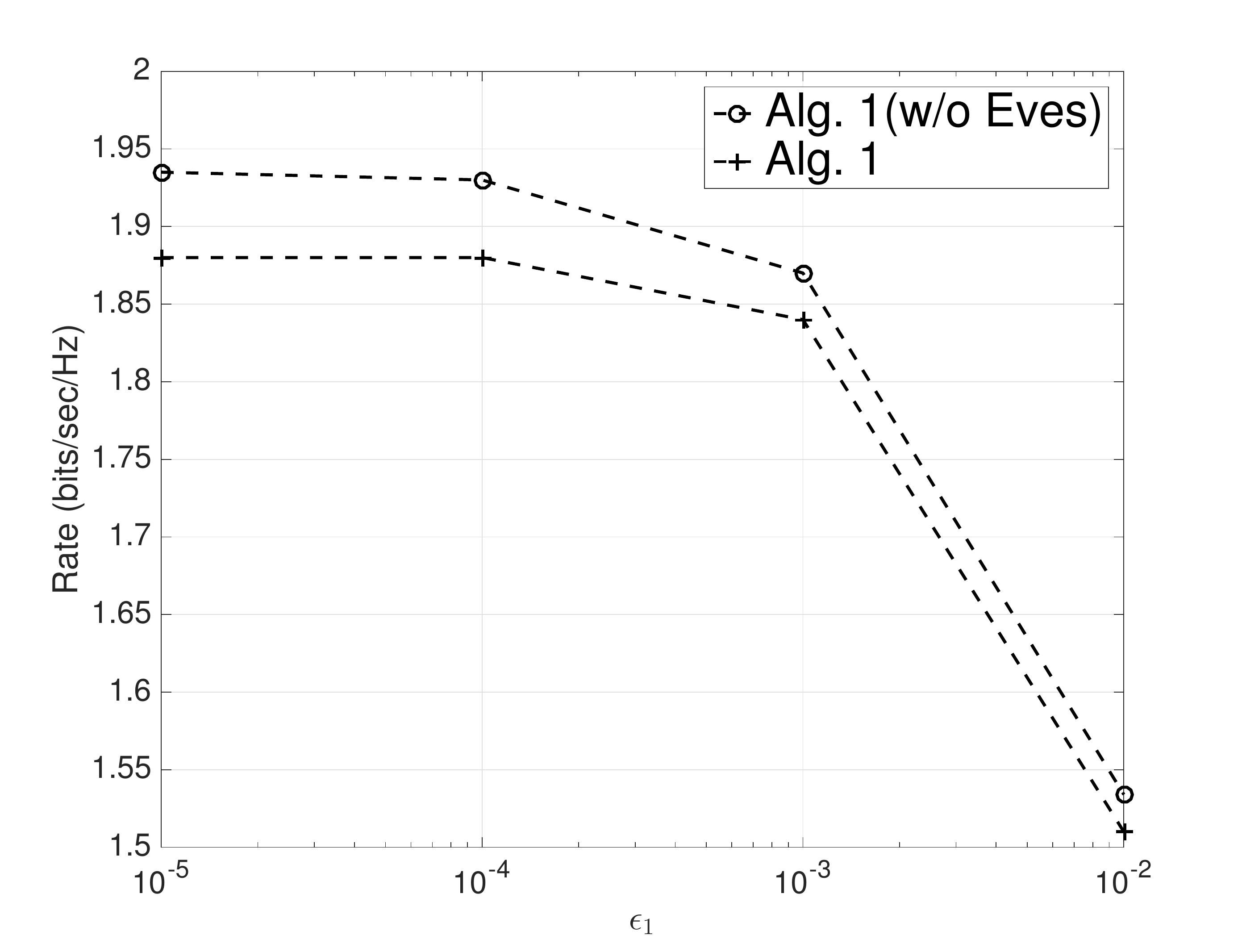}
  \caption{Robust secrecy rate (with eavesdroppers) and normal rate (without eavesdroppers) versus different levels of channel uncertainty $\epsilon_1$ for fixed energy harvesting threshold $e^\text{min} = -20$ dBm and BS antennas $M=5$.}
  \label{fig:Ue1}
  \end{minipage}
\end{figure*}
\else
\fi

Fig. \ref{fig:Ue0} plots the worst secrecy  and normal rates (in the absence of eavesdroppers) versus different levels of channel uncertainty in neighboring users' channels and the eavesdroppers' channels $\epsilon_0 = \{10^{-5},\hdots,10^{-2}\}$ for fixed $\epsilon_1 = 10^{-3}$, while Fig. \ref{fig:Ue1} plots such rates versus different levels of channel uncertainty in serving BS users' channels $\epsilon_0 = \{10^{-5},\hdots,10^{-2}\}$ for fixed $\epsilon_0 = 0.005$. We set energy harvesting threshold $e^\text{min} = -20$ dBm and number of BS antennas $M=5$. We can observe from Figs. \ref{fig:Ue0} and \ref{fig:Ue1} that the optimized rate is almost un-affected for low channel uncertainties $\{10^{-5},\hdots,10^{-3}\}$, and is slightly reduced if channel uncertainty is increased to the level of $10^{-2}$. This advocates the robustness of  Algorithm \ref{alg:1}. Even for wide range of values of channel uncertainty $\epsilon_0$, the optimized secrecy rate by  Algorithm \ref{alg:1} is quite close to that achieved by the modified algorithm, which assumes absence of eavesdroppers in Algorithm \ref{alg:1}.

Fig. \ref{fig:TS_PS} compares the secrecy rate performance of the proposed transmit TS-based system with that of the power splitting (PS)-based system \cite{Nasir-16-TSP-A} under the perfect channel state information ($\epsilon_0 = 0$, $\epsilon_1 = 0$). {\color{black}For the PS-based receiver in \cite{Nasir-16-TSP-A}, we set the ID circuit noise variance $\sigc$ to be $-90$ dBm and the antenna noise variance $\siga = -90$ dBm. Thus, for fair comparison in Fig. \ref{fig:TS_PS}, we add $\sigc$ to $\siga$, i.e., $\siga = -87$ dBm, for plotting the result for our proposed TS-based Algorithm \ref{alg:1}.} Fig. \ref{fig:TS_PS} plots the worst secrecy rate versus number of antennas $M$ for fixed energy harvesting threshold $e^\text{min} = -20$ dBm. We can clearly observe a gain of around $0.5$ bits/sec/Hz in the achieved secrecy rate of  Algorithm \ref{alg:1} compared to that of the algorithm in \cite{Nasir-16-TSP-A}. Note that the proposed TS-based system not only enjoys the performance gain, but also promises implementation simplicity. The average number of iterations required for convergence are almost same for both Algorithm \ref{alg:1} and the algorithm in \cite{Nasir-16-TSP-A}. However, the proposed TS-based system model has a two-fold advantage; it not only enjoys the performance gain, but also promises implementation simplicity, as motivated in the Introduction in Section \ref{sec:int}.

{\color{black} The computational complexity of the proposed Algorithm \ref{alg:1} is  $\mathcal{O} \left( i_\text{A1}  (MK (N+N_1) +1)^3  \left( 7 KN  \right. \right.$  $\left. \left. + (K+2) + 3 K N_1 \right) \right)$ \cite{Peaucelle-02-A}. Here, $i_\text{A1} = 22.5$ is the average number of times that \eqref{eq:P4} is solved by Algorithm \ref{alg:1}. Table \ref{tab:ca} shows the average number of iterations, scalar variables, and linear and quadratic constraints required to solve per iteration by the proposed Algorithm \ref{alg:1} and the PS-based algorithm in \cite{Nasir-16-TSP-A}. We can observe that though the PS-based algorithm in \cite{Nasir-16-TSP-A} requires the solution of fewer quadratic and linear constraints,  it is not practically easy to implement a variable range power splitter. Thus, the proposed TS-based Algorithm \ref{alg:1} provides a practical solution to secure and robust beamforming.}

\begin{table*}[t]
{\color{black}
\vspace{1cm}
\caption{Complexity analysis for the proposed TS-based Alg.~\ref{alg:1} and PS-based algorithm in \cite{Nasir-16-TSP-A} for general $M$, $N$, $K$, and specific $M=4$, $N=4$, $K=3$ cases.} \centering
\begin{tabular}{|c||c|c|c|c|} \hline
Algorithms    &  avg. $\#$ iter & scalar variables & linear constraints & quadratic constraints  \\ \hline \hline
Algorithm~\ref{alg:1} & 22.5 & $ MK (N+N_1) + 1 = 73$  & $3KN + KN_1 + (K+1) = 46$ &  $4KN + 2K N_1 +1=61$ \\ \hline
PS Algorithm \cite{Nasir-16-TSP-A} & 16 &  $MK (N+1) + KN_1 = 66$ &  $K+1=4$ & $KN+3KN_1 = 30$  \\ \hline
 \end{tabular}\label{tab:ca}}
\end{table*}

   \ifCLASSOPTIONpeerreview
\begin{figure*}[t]
    \centering
    \begin{minipage}[h]{0.45\textwidth}
    \centering
    \includegraphics[width=1.01 \textwidth]{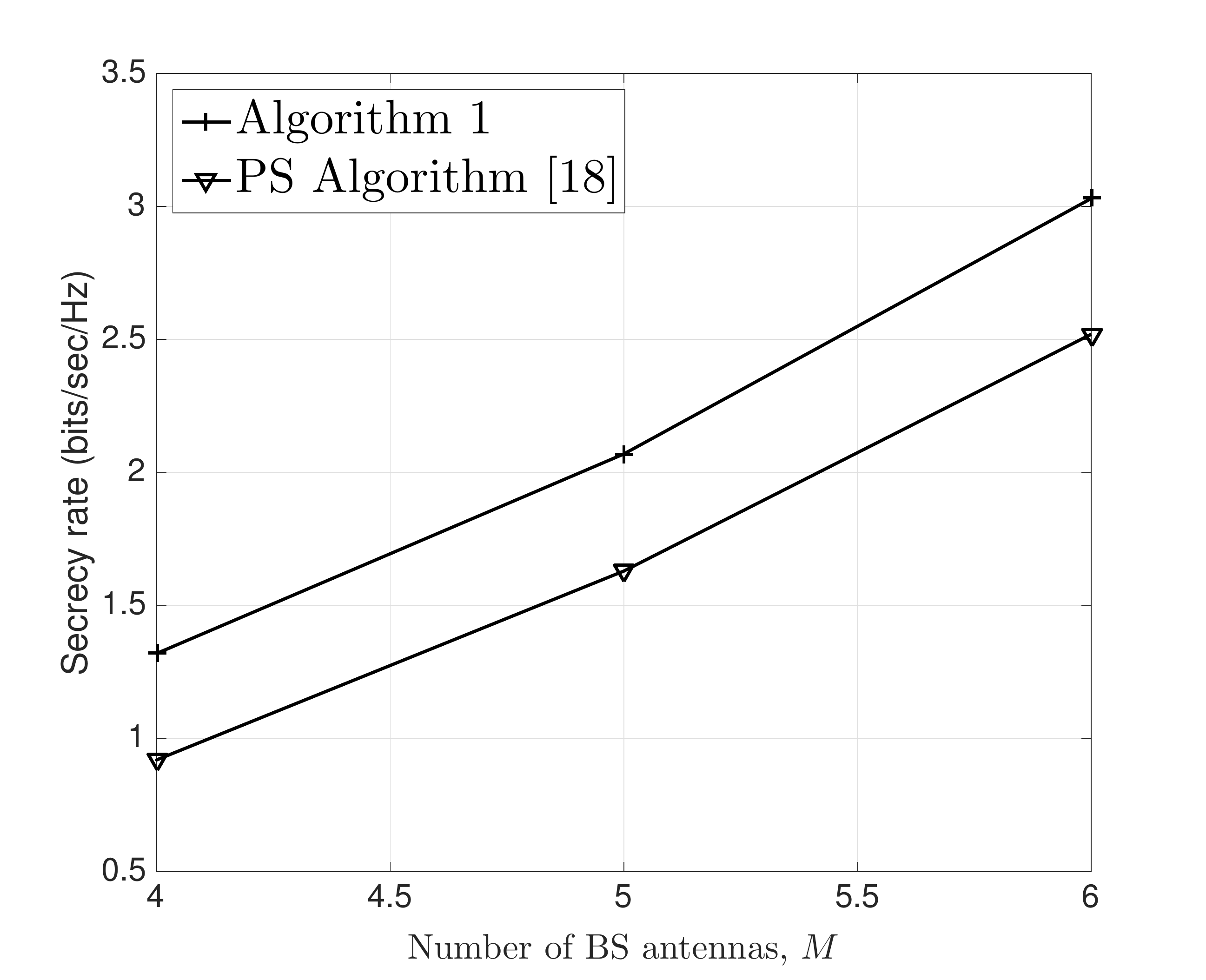}
  \caption{Comparison of proposed secrecy rate TS-based Algorithm \ref{alg:1} with the existing PS-based algorithm \cite{Nasir-16-TSP-A}  for fixed energy harvesting threshold $e^\text{min} = -20$ dBm and perfect channel estimation $\epsilon_0 = \epsilon_1 = 0$.}
  \label{fig:TS_PS}
  \end{minipage}
    \hspace{0.3cm}
    \begin{minipage}[h]{0.48\textwidth}
    \centering
    \includegraphics[width=1.01 \textwidth]{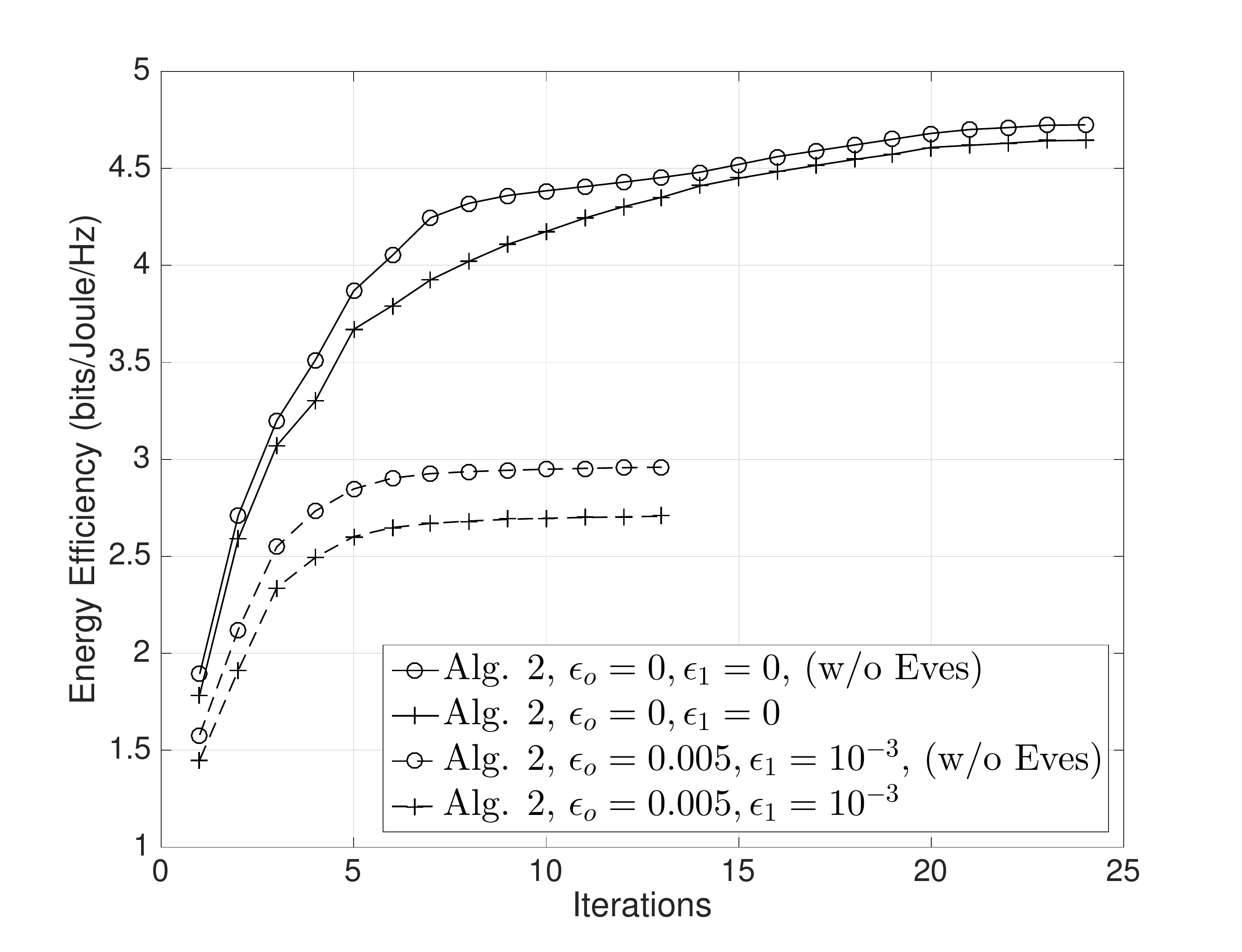}
  \caption{Convergence of Algorithm \ref{alg:2} for $M=5$ and $e^{\min} = -20$ dBm.}
  \label{fig:conv_EE}
  \end{minipage}
\end{figure*}
\else
\fi

Finally, the performance of our proposed SEE Algorithm \ref{alg:2} is evaluated.
Fig. \ref{fig:conv_EE} shows the convergence of proposed Algorithm \ref{alg:2} for $M=5$ antennas at the BS and  energy harvesting threshold $e^{\min} = -20$ dBm. We can see that for some fixed channel, whether we assume perfect channel estimation $\epsilon_0 = 0$, $\epsilon_1 = 0$ or assume some channel uncertainty $\epsilon_0 = 0.005$, $\epsilon_0 = 10^{-3}$, Algorithm \ref{alg:2} converges within $20-25$ iterations. On average, Algorithm \ref{alg:2} requires approximately $18.5$ iterations for convergence.

\ifCLASSOPTIONpeerreview
\begin{figure*}[t]
    \centering
    \begin{minipage}[h]{0.48\textwidth}
    \centering
    \includegraphics[width=1.01 \textwidth]{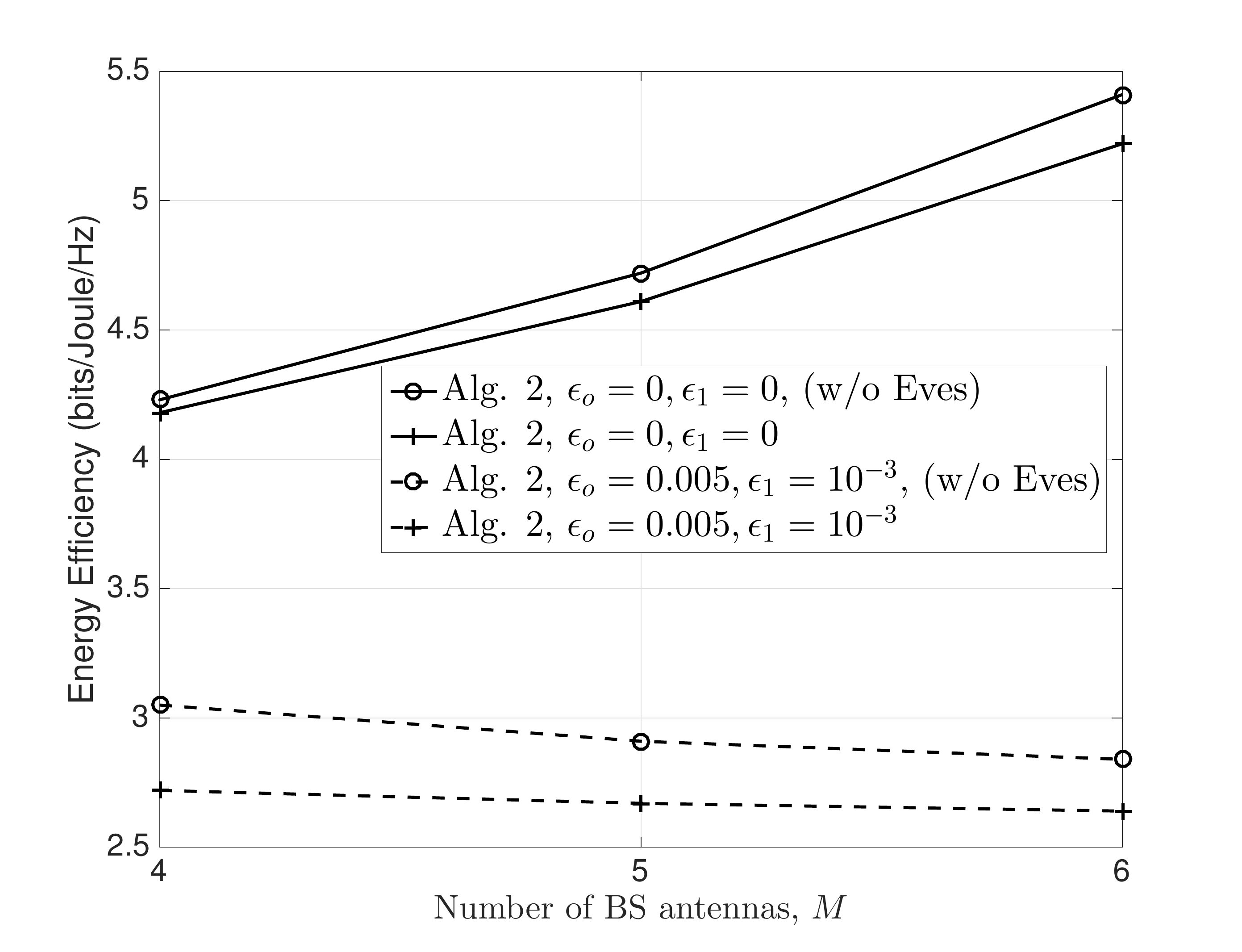}
  \caption{Robust secrecy energy efficiency and normal energy efficiency in the presence and absence of eavesdroppers, respectively, for varying number of antennas $M$ with fixed EH threshold $e^\text{min} = -20$ dBm.}
  \label{fig:EEM}
  \end{minipage}
    \hspace{0.3cm}
    \begin{minipage}[h]{0.48\textwidth}
    \centering
    \includegraphics[width=1.01 \textwidth]{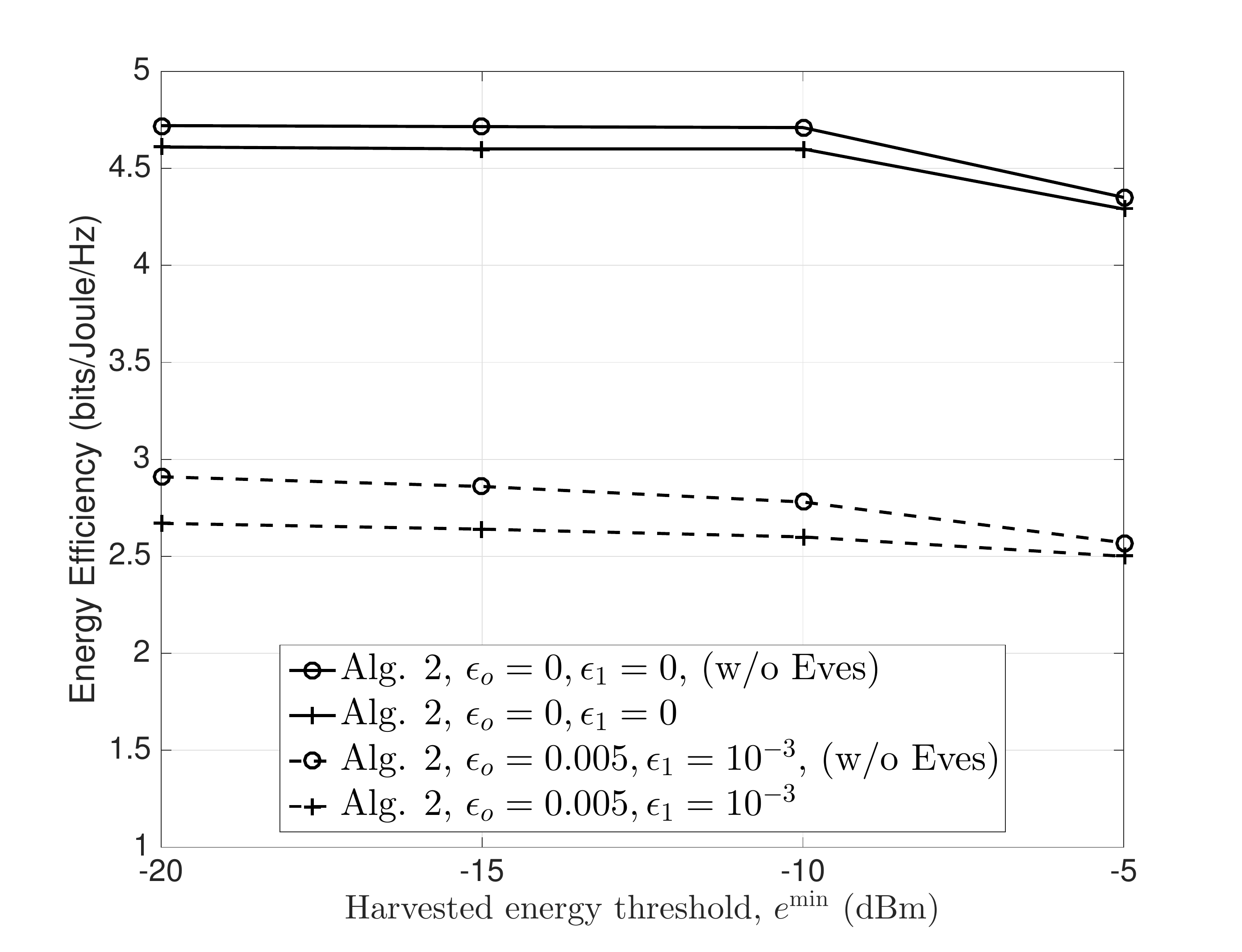}
  \caption{Robust secrecy energy efficiency and normal energy efficiency in the presence and absence of eavesdroppers, respectively, for different values of EH threshold $e^\text{min}$ with fixed number of BS antennas $M=5$.}
  \label{fig:EEH}
  \end{minipage}
\end{figure*}
\else
\fi

Figs. \ref{fig:EEM} and \ref{fig:EEH} plot the secrecy energy efficiency and normal energy efficiency (in the absence of eavesdroppers) for both perfect channel estimation $\epsilon_0 = 0$, $\epsilon_1 = 0$ and with the presence of channel uncertainty of $\epsilon_0 = 0.005$, $\epsilon_1 = 10^{-3}$. Here, the achievable SEE for  Algorithm \ref{alg:2} is compared with the normal EE assuming no eavesdroppers, that is obtained by solving the optimization problem
 \eqref{eP1} with $f_{k,n}^2(\mathbf{x},\eta)\equiv 0$. The dashed curves in Figs. \ref{fig:EEM} and \ref{fig:EEH}  refer to the presence of channel uncertainties $\epsilon_0 = 0.005$, $\epsilon_1 = 10^{-3}$, while solid line curves refer to the absence of channel uncertainty $\epsilon_0 =  \epsilon_1 = 0$. We can observe from Figs. \ref{fig:EEM} and \ref{fig:EEH} that the optimized EE by the proposed Algorithm \ref{alg:2} is quite close to that achieved by the modified algorithm, which assumes absence of eavesdroppers in Algorithm \ref{alg:2}. Finally, we observe from Fig. \ref{fig:EEM}  that for perfect channel estimation, the optimized EE increases by increasing the number of antennas, as per expectation, however, in the presence of channel uncertainties, the EE decreases with the increase in the number of antennas. In order to investigate this, we have figured out the numerator and denominator of EE function separately in the next two figures.

 \ifCLASSOPTIONpeerreview	
\begin{figure*}[t]
    \centering
    \begin{minipage}[h]{0.48\textwidth}
    \centering
    \includegraphics[width=1.01 \textwidth]{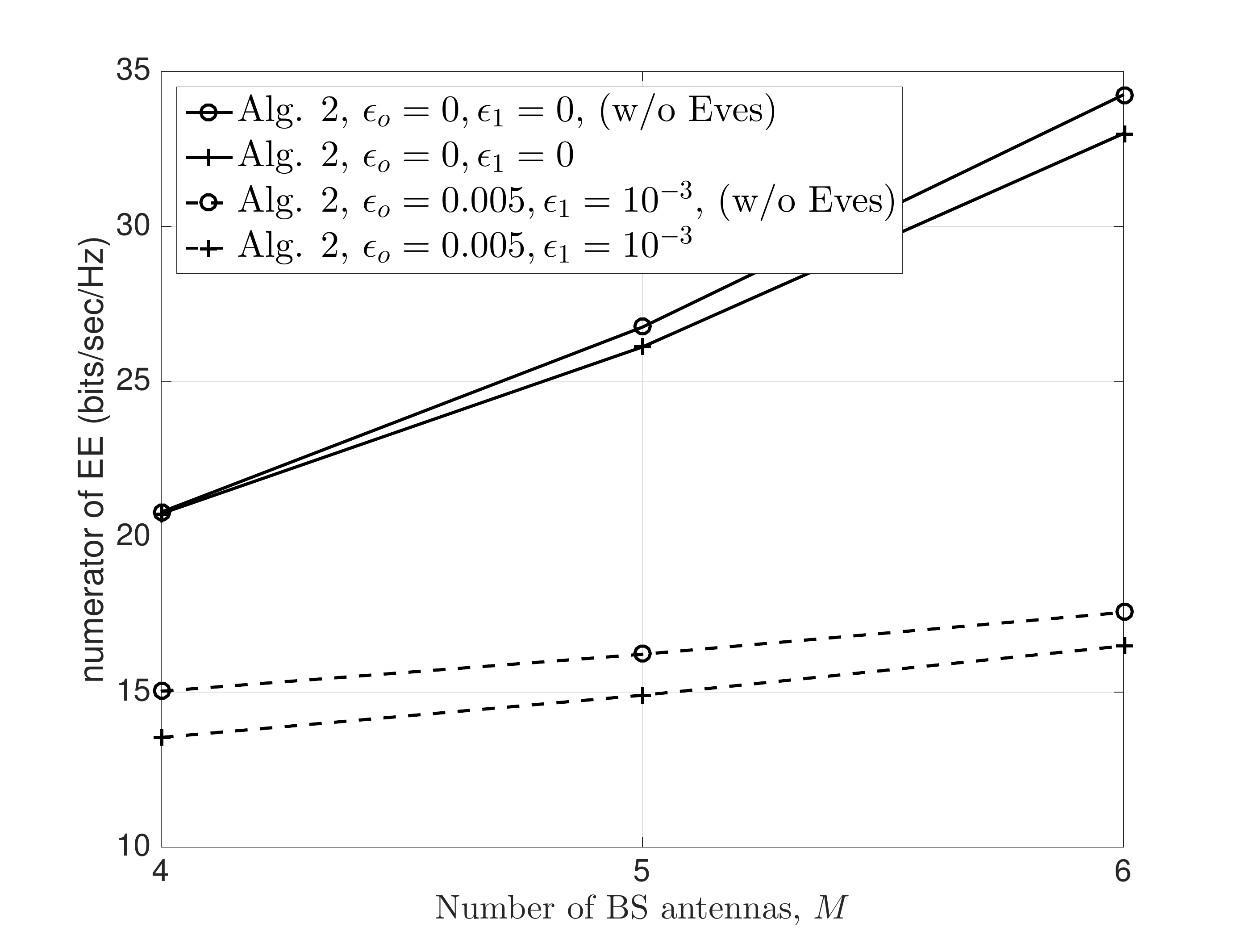}
  \caption{Numerator of energy efficiency (sum-rate of the worst cell) in the presence of eavesdroppers for varying number of antennas $M$ with fixed EH threshold $e^\text{min} = -20$ dBm.}
  \label{fig:EEMn}
  \end{minipage}
    \hspace{0.3cm}
    \begin{minipage}[h]{0.48\textwidth}
    \centering
    \includegraphics[width=1.01 \textwidth]{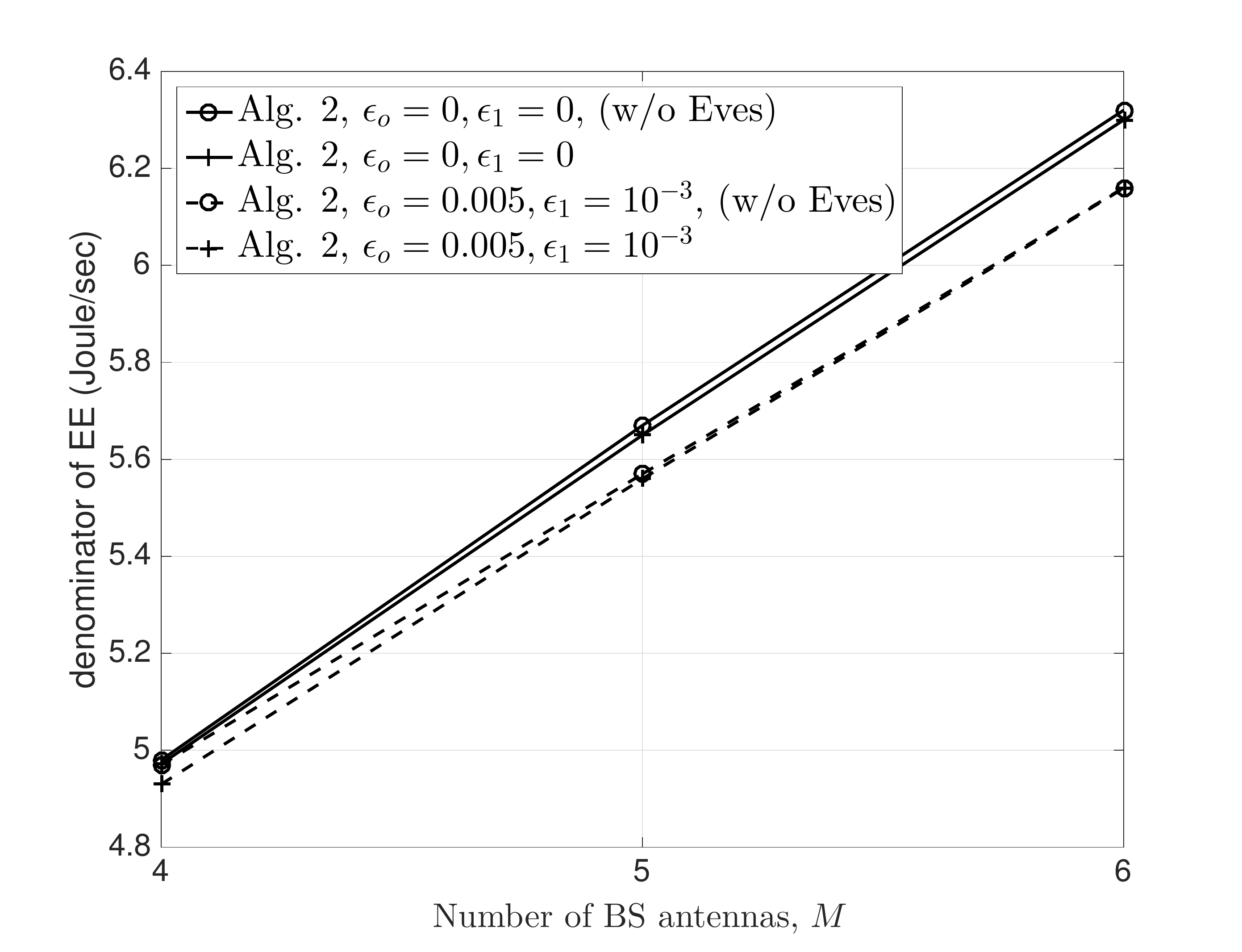}
  \caption{Denominator of energy efficiency (total power of the worst cell's BS) in the presence of eavesdroppers for varying number of antennas $M$ with fixed EH threshold $e^\text{min} = -20$ dBm.}
  \label{fig:EEMd}
  \end{minipage}
\end{figure*}
\else
\fi

In Figs. \ref{fig:EEMn} and \ref{fig:EEMd}, we plot the numerator and denominator of the EE function, \eqref{eP1a}, respectively, where numerator of EE corresponds to the sum-rate of the worst cell and denominator of EE corresponds to the total power, $\frac{1}{\xi}g_k(\mathbf{x}_k,\eta)+MP_A+P_c$, of the worst cell's BS. First, we can obersve from Fig. \ref{fig:EEMd} that denominator of EE function increases with the increase in the number of antennas. This This is because increase in the number of antennas increases the non-transmission power $M P_A$ in the denominator of EE function, \eqref{eP1a}. Next, we can observe from Fig. \ref{fig:EEMn} that though the numerator part of EE function also increases with the increase in the number of antennas, however, the numerator part increases swiftly for the prefect channel estimation case (solid lines) when compared to its slow increase in the presence of channel uncertainties (dashed lines). This results in slight decrease in the EE with the increase in the number of antennas in the presence of erroneous estimates, as shown in Fig. \ref{fig:EEM}.

\section{Conclusions}\label{sec:conclude}
Considering simple and efficient transmit TS approach to ensure wireless energy harvesting and information decoding in a dense multi-cell network, we have proposed robust secrecy rate and secrecy energy efficiency maximization algorithms in the presence of multi-antenna eavesdroppers and channel estimation errors. Our robust optimization algorithm jointly designs for transmit energy and information beamformers at the BSs and the transmit TS ratio with the objective of maximizing the worst-case user secrecy rate under BS transmit power and UE minimum harvested energy constraints. The problem is very challenging due to nonconvex objective and numerous  non-convex constraints. We have solved it by a new robust path-following algorithm, which involves one simple convex quadratic program in each iteration. We have also extended our algorithm to solve for worst cell secrecy EE maximization problem under secrecy rate quality-of-service constraints, which is further complex due to additional function of optimization variables in the denominator of secrecy rate function. Our numerical results confirm the merits of the proposed algorithms as their performance is quite close to that of the case where there are no eavesdroppers. Moreover, the proposed algorithm not only outperforms the existing algorithm that models PS based receiver but also the proposed transmit TS based model is implementation-wise quite simple than the PS-based model. Finally, we would like to hint towards an open future research direction from this work, in which the proposed algorithms could be modified to accommodate only channel distribution knowledge about the eavesdroppers. {\color{black} The rate-energy trade-off between transmit-TS and receive-PS approaches could also be considered in future research.}

\appendices
\numberwithin{equation}{section}

\section{\vspace{-0pt}Proof of Theorem 1 }\label{app:A}
{\color{black}We first prove \eqref{eq:f1_kappa} by using
the following inequality for all $x>0$, $\bar{x}>0$, $t>0$ and $\bar{t}>0$
\begin{align}\label{eq:ln_ineq}
\ds\frac{\ln(1+1/x)}{t}&\geq f(\bar{x},\bar{t})+\la \nabla f(\bar{x},\bar{t}), (x,t)-(\bar{x},\bar{t})\ra\nonumber\\
&=2\ds\frac{\ln(1+1/\bar{x})}{\bar{t}}+\frac{1}{\bar{t}(\bar{x}+1)}-\frac{x}{(\bar{x}+1)\bar{x}\bar{t}}-
\frac{\ln(1+1/\bar{x})}{\bar{t}^2}t.
\end{align}
which follows from the convexity of function $\frac{\ln(1+1/x)}{t}$.}

By subsituting $1/x\rightarrow x$ and $1/\bar{x}\rightarrow \bar{x}$ in \eqref{eq:ln_ineq}, we have:
\begin{align}
\ds\frac{\ln (1+x)}{t}&\geq a-\frac{b}{x}-ct,\label{c4}
\end{align}
where $a=2\frac{\ln(1+\bar{x})}{\bar{t}}+\frac{\bar{x}}{\bar{t}(\bar{x}+1)}>0,
b=\frac{\bar{x}^2}{\bar{t}(\bar{x}+1)}>0, c=\frac{\ln(1+\bar{x})}{\bar{t}^2}>0$. From that,
\begin{align}\label{c5}
\frac{1}{\mu}\ln\left(1+\ds\frac{\left(\Re\{\mathbf{h}_{k,k,n}^H \mathbf{x}^I_{k,n} \}\right)^2 - \epsilon_{k,k,n} \| \mathbf{x}_{k,n}^I \|^2}{\varphi_{k,n}(\mathbf{x}^I)}\right) &\geq
a^{(\ell)}-b^{(\ell)}\ds\frac{\varphi_{k,n}(\mathbf{x}^I)}{\left(\Re\left\{\mathbf{h}_{k,k,n}^H \mathbf{x}^I_{k,n}\right\}\right)^2 - \epsilon_{k,k,n} \| \mathbf{x}_{k,n}^I \|^2 }-c^{(\ell)}\mu
\end{align}
where $a^{(\ell)}$, $b^{(\ell)}$, $c^{(\ell)}$, and $d^{(\ell)}$ are defined in \eqref{c6}. Now, using
$(\Re\{\mathbf{h}^H_{k,k,n}\mathbf{x}^I_{k,n}\})^2\geq \psi_{k,n}(\mathbf{x}^I_{k,n}) $
with $\psi_{k,n}(\mathbf{x}^I_{k,n}) \ge 0$ defined in
(\ref{ap1}), together with (\ref{c5}) leads to
\begin{align}\label{c62}
\frac{1}{\mu}\ln\left(1+\ds\frac{\left(\Re\left\{\mathbf{h}_{k,k,n}^H \mathbf{x}^I_{k,n}\right\}\right)^2 - \epsilon_{k,k,n} \| \mathbf{x}_{k,n}^I \|^2}{\varphi_{k,n}(\mathbf{x}^I)}\right) &\geq
  a^{(\ell)}-b^{(\ell)}\ds\frac{\varphi_{k,n}(\mathbf{x}^I)}{\nu_{k,n}(\mathbf{x}^I_{k,n}) }-c^{(\ell)}\mu \notag \\ &\triangleq \bar{f}^{1,(\ell)}_{k,n}(\mathbf{x}^I,\mu)
\end{align}
for $0 \le \nu_{k,n}(\mathbf{x}^I_{k,n}) \le \psi_{k,n}(\mathbf{x}^I_{k,n})-\epsilon_0||\mathbf{x}^I_{k,n}||^2, \ \forall k\in\clK, n\in\clN_k$.
The function $\bar{f}^{1,(\ell)}_{k,n}(\mathbf{x}^I,\mu)$ is concave on (\ref{c7}).

Next, \eqref{eq:f2_kappa} follows from the following inequality
\[
\ln (1+t) \leq \ln(1+t') +(t-t')/(1+t')\ \forall t\geq 0, t'\geq 0,
\]
which is a consequence of the concavity of function $\ln(1+t)$.

{\color{black}Now, it remains to prove  \eqref{eq:qk_beta}.} By substituting $\bar{q}_{k,n}(\mathbf{x},\mu)$, defined in \eqref{eq:qk_beta}, into the constraint \eqref{etakn1}, we have:
\begin{eqnarray}
\ds\frac{{\color{black}\sqrt{\beta_{k,n}}}}{\mu-1}+\frac{1}{\mu-1}\left(\ds\sum_{\bn\in\mathcal{N}_{k} \setminus \{n\}} \epsilon_{k,k} \|\mathbf{x}^I_{k,\bn} \|^2+ \ds\sum_{\bk\in\mathcal{K}\setminus\{k\}} \sum_{\bn\in\mathcal{N}_{\bk}} \epsilon_{\bk,k} \|\mathbf{x}^I_{\bk,\bn} \|^2 \right)
+ \ds\sum_{\bk\in\mathcal{K}} \sum_{\bn\in\mathcal{N}_{1,k}} \epsilon_{\bk,k} \|\mathbf{x}^E_{\bk,\bn} \|^2&\leq&\nonumber\\
\ds  \ds\sum_{\bk\in\mathcal{K}} \sum_{\bn\in\mathcal{N}_{1,k}} \|\boldsymbol{\mathcal{H}}_{\bk,k}^H \mathbf{x}^E_{\bk,\bn} \|^2
 +\frac{1}{\mu-1}\left(\ds\sum_{\bn\in\mathcal{N}_{k} \setminus \{n\}} \|\boldsymbol{\mathcal{H}}_{k,k}^H \mathbf{x}^I_{k,\bn} \|^2 +\ds\sum_{\bk\in\mathcal{K}\setminus\{k\}} \sum_{\bn\in\mathcal{N}_{\bk}} \|\boldsymbol{\mathcal{H}}_{\bk,k}^H \mathbf{x}^I_{\bk,\bn} \|^2\right)&&\nonumber\\
  + (1+\frac{1}{\mu-1}) N_\text{ev}\siga,&&\label{etakn2}
\end{eqnarray}
where right hand side of (\ref{etakn2}) is convex which can be linearized for inner approximation by using \cite{Nasir-16-CL-A}
\begin{align}\label{eq:append_app_12}
\frac{\| \mathbf{x}\|^2}{y} \geq \frac{2\Re\left\{(\mathbf{x}^{(\ell)})^H\mathbf{x}\right\}}{y^{(\ell)}}
- \frac{\| \mathbf{x}^{(\ell)}\|^2 y}{\left(y^{(\ell)}\right)^2}, \quad \forall \mathbf{x}\in\mathbb{C}^N,
\mathbf{x}^{(\ell)}\in\mathbb{C}^N, y > 0, y^{(\ell)} > 0,
\end{align},
\begin{align}\label{eq:append_app_1}
\| \mathbf{x}\|^2 \geq 2\Re\left\{(\mathbf{x}^{(\ell)})^H\mathbf{x}\right\}
-\| \mathbf{x}^{(\ell)}\|^2, \quad \forall \mathbf{x}\in\mathbb{C}^N,
\mathbf{x}^{(\ell)}\in\mathbb{C}^N
\end{align}
and
\begin{align}\label{eq:append_app_2}
\frac{1}{\mu-1} \ge \frac{2}{\mu^{(\ell)} -1} - \frac{\mu-1}{(\mu^{(\ell)} - 1)^2}.
\end{align}
The first term in the left hand side of (\ref{etakn2}) is nonconvex, which is convexified by using the fact that $\sqrt{xy}$ is concave in $x$ and $y$, i.e., $ \sqrt{xy} \le \frac{\sqrt{s^{(\ell)}} y}{2 \sqrt{y^{(\ell)}} } + \frac{\sqrt{y^{(\ell)}} x}{2 \sqrt{s^{(\ell)} }}$. Thus, $\frac{{\color{black}\sqrt{\beta_{k,n}}}}{\mu-1}$ is approximated as
\begin{equation}\label{etakn3}
{\color{black}
\ds\frac{\sqrt{\beta_{k,n}}}{\mu-1}\leq \frac{1}{2} \left( \frac{\beta_{k,n}}{\sqrt{\beta_{k,n}^{(\ell)}}(\mu^{(\ell)}-1)}
+\frac{\sqrt{\beta_{k,n}^{(\ell)}}(\mu^{(\ell)}-1)}{(\mu-1)^2} \right)}
\end{equation}
Thus, using \eqref{eq:append_app_12}, \eqref{eq:append_app_1}, \eqref{eq:append_app_2}, and \eqref{etakn3} in \eqref{etakn2}, we can get the approximation \eqref{etakn4}.


\end{document}